\documentclass{article}
\usepackage[utf8]{inputenc}
\usepackage{algorithm}
\usepackage{algpseudocode}
\usepackage{authblk}
\usepackage[top=1in, bottom=1in, left=1in, right=1in]{geometry}
\usepackage{tikz}
\usetikzlibrary{shapes,arrows,matrix,patterns,positioning}
\usepackage{color}
\usepackage{graphicx}
\usepackage{amssymb}
\usepackage{epstopdf}

\usepackage{float}
\usepackage{amsmath,amsthm,bm,color,epsfig,enumerate,caption}
\usepackage{hyperref}
\usepackage{booktabs}
\usepackage{multirow}
\usepackage{array}
\newtheorem{theorem}{Theorem}

\title{Spectral clustering via orthogonalization-free methods}
\author[1]{Qiyuan Pang}
\author[2]{Haizhao Yang \thanks{corresponding author, email: hzyang@umd.edu}}
\affil[1]{Department of Mathematics, Purdue University}
\affil[2]{Department of Mathematics and Department of Computer Science\\ University of Maryland College Park}

\date{\vspace{-5ex}}

\begin{document}

\maketitle

\begin{abstract}

While orthogonalization exists in current dimensionality reduction methods in spectral clustering on undirected graphs, it does not scale in parallel computing environments. We propose four orthogonalization-free methods for spectral clustering. Our methods optimize one of two objective functions with no spurious local minima. In theory, two methods converge to features isomorphic to the eigenvectors corresponding to the smallest eigenvalues of the symmetric normalized Laplacian. The other two converge to features isomorphic to weighted eigenvectors weighting by the square roots of eigenvalues. We provide numerical evidence on the synthetic graphs from the IEEE HPEC Graph Challenge to demonstrate the effectiveness of the orthogonalization-free methods. Numerical results on the streaming graphs show that the orthogonalization-free methods are competitive in the streaming graph scenario since they can take full advantage of the computed features of previous graphs and converge fast. Our methods are also more scalable in parallel computing environments because orthogonalization is unnecessary. Numerical results are provided to demonstrate the scalability of our methods. Consequently, our methods have advantages over other dimensionality reduction methods when handling spectral clustering for large streaming graphs.

\end{abstract}


\section{Introduction}
Spectral clustering uses the spectrum of the (normalized) Laplacian matrix of a graph to perform dimensionality reduction before clustering in low dimensions. The basic algorithm is summarized as \cite{ng2001spectral, weiss1999segmentation}:

\begin{itemize}
    \item[] 1. Calculate the symmetric normalized Laplacian $\mathbf{L}$ of a undirected graph $\mathcal{G}$ with $N$ graph nodes;
    \item[] 2. Compute the first $k$ orthonormal eigenvectors $\mathbf{U}_{k}$ corresponding to the smallest $k$ eigenvalues of $\mathbf{L}$.
    \item[] 3. Normalize each row of $\mathbf{U}_{k}$ and use the resulting matrix as the feature matrix where the $l$-th row defines the features of graph node $l$.
    \item[] 4. Cluster the graph based on the features using clustering methods like Euclidean distance K-means clustering.
\end{itemize}

We intend to limit the discussion in this paper to Euclidean space. Given that spectral clustering is primarily applied within this context, we believe this restriction does not adversely affect the paper's scope. There are different versions of spectral clustering based on unnormalized Laplacian \cite{von2007tutorial} and random walk normalized Laplacian \cite{shi2000normalized}. And normalizing rows of $\mathbf{U}_{k}$ is optional \cite{von2007tutorial}. In this paper, we focus on the version of spectral clustering summarized above. We first introduce the symmetric Laplacian $\mathbf{L}$ as follows,
\begin{equation}
    \mathbf{L} = \mathbf{I} - \mathbf{D}^{-1/2}\mathbf{S}\mathbf{D}^{-1/2},
\end{equation}
where $\mathbf{I}$ is the identity matrix, $\mathbf{S}$ is the similarity matrix with $\mathbf{S}_{ij} = 1$ if nodes $i$ and $j$ are connected otherwise $0$, and $\mathbf{D}$ is the diagonal degree matrix with $\mathbf{D}_{ii} = \sum_{j}\mathbf{S}_{ij}$. The smallest eigenvalue of $\mathbf{L}$ is always $0$ and the largest eigenvalue is $2$ \cite{von2007tutorial}. For more details of the symmetric normalized Laplacian $\mathbf{L}$, please refer to \cite{von2007tutorial}. Hereafter, for the sake of simplicity, we refer to $\mathbf{L}$ as the normalized Laplacian if no additional statements are provided. In the rest of this paper, $\mathbf{U_{k}}$ is orthonormal vectors.

Most works in the literature and software for the dimensionality reduction in spectral clustering fall into two categories: computing the eigenvector matrix $\mathbf{U}_{k}$ directly or approximating $\mathbf{U}_{k}$. Works including \cite{belkin2003laplacian, ng2001spectral, zelnik2004self} and software like scikit-learn use eigensolvers like LOBPCG \cite{knyazev2001toward} with algebraic multigrid (AMG) preconditioning and ARPACK \cite{lehoucq1998arpack} to compute $\mathbf{U}_{k}$. LOBPCG is also used without preconditioning because AMG preconditioning is not always effective for all graphs \cite{zhuzhunashvili2017preconditioned}. Recently, a block Chebyshev-Davidson method \cite{bchdav4clustering2022} is also used to compute $\mathbf{U}_{k}$ because the well-known analytic bounds of the normalized Laplacian $\mathbf{L}$ could accelerate the convergence of the method. Parallel LOBPCG \cite{naumov2016parallel}, ARPACK \cite{chen2010parallel, huo2020designing}, and block Chebyshev-Davidson method \cite{bchdav4clustering2022} are also used in parallel settings to enable spectral clustering for large graphs. Due to the rapid growth of machine learning and its various applications \cite{ZHANG2024105090, zhang2023effect, Weng2024, Weng202404, dan2024evaluation} in recent years, many non-spectral type methods,  for instance, kernel-based methods \cite{xu2022multi, xu2022data}, have been developed for clustering. However, they are not in the scope of this paper because we concentrate on spectral-type methods.

Instead of computing $\mathbf{U}_{k}$ directly, other works concentrate on approximating $\mathbf{U}_{k}$, e.g., using Power Iteration (PI) \cite{lin2010power, boutsidis2015spectral, ye2016fuse} and Graph Signal Filtering/Processing (GSF) \cite{paratte2016fast, ramasamy2015compressive, tremblay2016compressive, li2019fast}. Power Iteration Clustering \cite{lin2010power} computes a one-dimensional feature using power iteration with early stopping. Boutsidis et al. \cite{boutsidis2015spectral} extend the work by applying the normalized similarity matrix to the power $p$ to $k$ Gaussian signals and then compute the left singular vectors of the $N \times k$ obtained matrix. Ye et al. \cite{ye2016fuse} propose FUSE, which first generates $m(m > k)$ pseudo-eigenvectors by merging the eigenvectors obtained by power iteration and then applies Independent Component Analysis to rotate the pseudo-eigenvectors to make them pairwise statistically independent. Parallel Power Iteration Clustering is developed in \cite{yan2013p}. Graph Signal Filtering estimates the first $k$ eigenvectors of any graph Laplacian via filtering i.i.d. $\mathcal{N}(0,1/d)$ signals where $d \geq k$ and orthogonalizing the filtered signals. GSF heavily relies on determining the $k$-th smallest eigenvalue, e.g., Tremblay et al. \cite{tremblay2016compressive} use the dichotomy and edge count techniques in \cite{di2016efficient}. Then, Paratte et al. \cite{paratte2016fast}, and Li et al. \cite{li2019fast} accelerate the estimation by assuming eigenvalues approximately satisfy local uniform distribution. Such estimations are usually expensive even under the uniform eigenvalues distribution assumption.

To facilitate efficient estimation of $\mathbf{U}_{k}$, we propose to approximate it by solving one of the two unconstrained optimization problems
\begin{equation}
    \min_{\mathbf{X}\in \mathbb{R}^{N\times k}} f_{1}(\mathbf{X}) =: \|\mathbf{A}+\mathbf{X} \mathbf{X}^{T}\|_{F}^2,
\end{equation}
and 
\begin{equation}
    \min_{\mathbf{X}\in \mathbb{R}^{N\times k}} f_{2}(\mathbf{X}) =: tr((2\mathbf{I}-\mathbf{X}^{T}\mathbf{X})\mathbf{X}^{T}\mathbf{A}\mathbf{X})
\end{equation}
where $\mathbf{A} = \mathbf{L} - 2\mathbf{I}$, $\|*\|_{F}$ denotes the Frobenius norm, and $tr(*)$ denotes the trace operation. Note that $\mathbf{A}$ and $\mathbf{L}$ have the same eigenvectors. Denote $\lambda_i$ as the $i$-th eigenvalue of $\mathbf{A}$ then $\lambda_i+2$ is the $i$-th eigenvalue of $\mathbf{L}$. Obviously, $\lambda_i \in [-2,0], \forall i$. Objective $f_{1}$ has been adopted to address the extreme eigenvalue problems via coordinate-wise methods
\cite{lei2016coordinate, li2020optimal, li2019coordinatewise}. Objective $f_{2}$ is widely known as the orbital minimization method (OMM) \cite{corsetti2014orbital, lu2017orbital, lu2017preconditioning, mauri1993orbital, ordejon1993unconstrained}, which is popular in the area of DFT. 
It is known that all local minima of these $f_{1}$ and $f_{2}$ are global minima \cite{li2019coordinatewise, liu2015efficient, lu2017orbital}. Though a local minimum might not be the desired eigenvectors in spectral clustering, it is isomorphic to the eigenvectors ($f_1$) or weighted eigenvectors ($f_2$). Therefore, a local minimum may be effective for spectral clustering. We will discuss this in the later sections. Unlike the methods based on ARPACK, LOBPCG, PI, and GSF,  solving the unconstrained optimization problems using gradient descent is orthogonalization-free, and we will refer to these methods as orthogonalization-free methods (OFM). To distinguish OFM with $f_{1}$ and $f_{2}$, we will denote OFM with objectives $f_{1}$ and $f_{2}$ as OFM-$f_{1}$ and OFM-$f_{2}$ respectively, throughout the paper. Recently, Gao et al. \cite{gao2022triangularized} proposed the so-called triangularized orthogonalization-free method (TriOFM) for addressing the two unconstrained optimization problems with additional assumptions that orthogonalization is not permitted and eigenvectors are sparse. Though the eigenvectors of any graph Laplacians are dense, we apply TriOFM for spectral clustering. We will refer to the methods with $f_1$ and $f_2$ as TriOFM-$f_{1}$ and TriOFM-$f_{2}$, respectively. The spectral clustering via one of the four orthogonalization-free methods is summarized as Algorithm \ref{algo:spclustering}. 

\begin{figure}
\begin{center}
    \begin{tabular}{cc}
        \includegraphics[scale=0.20]{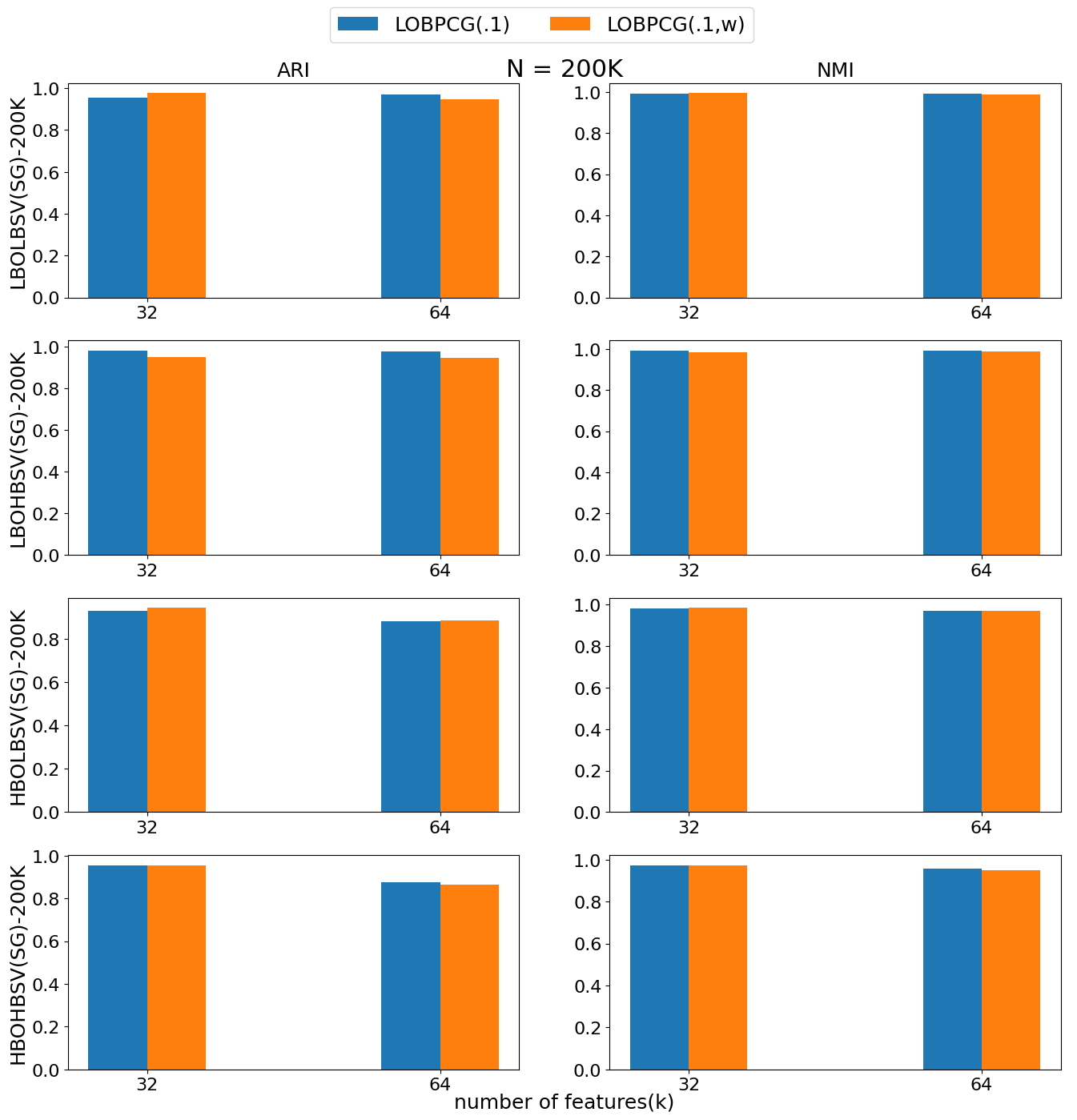} & 
        \includegraphics[scale=0.20]{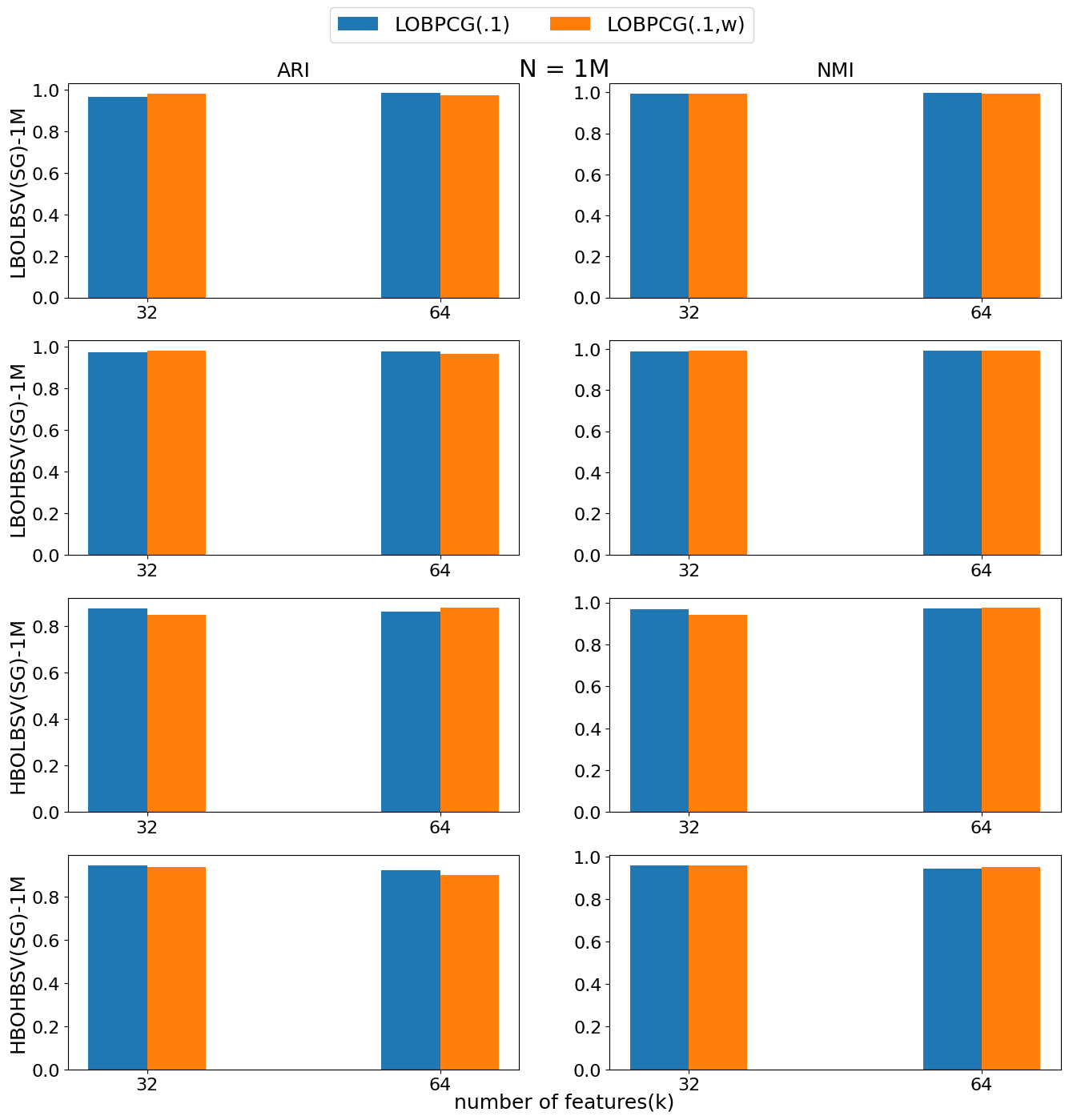}
    \end{tabular}
\end{center}
\caption{Clustering performance (ARI $\&$ NMI) based on eigenvectors $\mathbf{U}_{k}$ and weighted eigenvectors $\mathbf{U}_{k} \sqrt{-\Lambda_k}$ where $\Lambda_k$ contains the smallest eigenvalues of $\mathbf{A}$, on clustering static graphs each with 200 thousand nodes (left) or 1 million nodes (right). Eigenvectors $\mathbf{U}_{k}$ are first computed by LOBPCG with $0.1$ tolerance (`LOBPCG(.1)'), and weighted eigenvectors $\mathbf{U}_{k} \sqrt{-\Lambda_k}$ are then evaluated (`LOBPCG(.1,w)').}
\label{fig:compfeatures}
\end{figure}

This paper will examine theoretical assurances and present numerical evidence demonstrating that these orthogonalization-free methods provide suitable approximations of the matrix $\mathbf{U}_{k}$ or effective features for spectral clustering. Theoretically, OFM-$f_2$ and TriOFM-$f_2$ converge to a solution point that is unitary with respect to the eigenvectors $\mathbf{U}_k$, indicating that the solution point is equivalent to $\mathbf{U}_k$ in Euclidean space. Therefore, they are comparable in spectral clustering within this context. Meanwhile, OFM-$f_1$ and TriOFM-$f_1$ yield a solution point unitary to the eigenvectors $\mathbf{U}_k$ weighted by $\sqrt{-\lambda_i}$ where $-2\leq \lambda_i \leq 0$. Consequently, OFM-$f_1$ and TriOFM-$f_1$ are effectively using $\mathbf{U}_k$ weighted by $\sqrt{-\lambda_i}$ instead of $\mathbf{U}_k$ in spectral clustering. Based on extensive numerical evidence (e.g., Figure \ref{fig:compfeatures}), we hypothesize that the weighted $\mathbf{U}_k$ are also competitive features in spectral clustering. To address large data challenges, we have developed scalable parallel implementations of each method. See Table \ref{tb:complexity} for complexity summary.

Before proceeding with the paper, we summarize our contributions as follows.
\begin{itemize}
    \item[] 1. We demonstrate that our orthogonalization-free methods show competitive performance in spectral clustering compared to other methods which require orthogonalization.
    \item[] 2. We provide massive numerical evidence to show that our methods are competitive in the streaming graph scenario because the methods could take full advantage of computed features for precious graphs and converge fast.
    \item[] 3. In parallel computing environments, our methods are more scalable than other methods and have advantages over other methods when handling spectral clustering for large streaming graphs.
\end{itemize}

We organize the rest of the paper as follows. Section 2 describes how the orthogonalization-free methods work for dimensionality reduction in spectral clustering. Section 3 provides implementation details of our methods' sequential and parallel versions. Section 4 analyzes the complexity of the methods. Numerical results in Section 5 demonstrate the effectiveness and scalability of the methods.

\begin{algorithm}
\caption{Spectral Clustering via Orthogonalization-Free Methods}
\label{algo:spclustering}
\begin{algorithmic}[1]
\State \textbf{Input:} an undirected graph with $N$ graph nodes.

\State Calculate the symmetric normalized Laplacian $\mathbf{L}$ of the undirected graph, and then construct $\mathbf{A} = \mathbf{L} - 2\mathbf{I}$.
\State Use one of the Orthogonalization-Free methods to compute a feature matrix $\mathbf{Y}_{k}$ of size $N \times k$ by optimizing objectives $f_1$ or $f_2$ on $\mathbf{A}$.
\State Normalize each row of the feature matrix $\mathbf{Y}_{k}$ and use the resulting matrix as the feature matrix where the $l$-th row defines the features of graph node $l$.
\State Cluster the graph based on the features using clustering methods like K-means.
\State \textbf{Output:} the cluster assignments of the graph nodes.
\end{algorithmic}

\end{algorithm}

\begin{table}[!htbp]
\caption{\textbf{Per-iteration complexity of parallel orthogonalization-free methods with the assumption $k \ll N$ and only one single thread used for each process or node. $nnz(\mathbf{A})$ denotes the number of nonzeros elements of a sparse matrix $\mathbf{A}$. $p$ is the number of processes or compute nodes.}}
\centering
\scalebox{1.0}{
\begin{tabular}{|c|c|c|}
\hline
Methods & Flops & Communication cost \\
\hline
OFM-$f_1$ & $\dfrac{nnz(\mathbf{A})4k + 12Nk^2 + 12Nk}{p}$ & $20\alpha \log p + \beta(\dfrac{8Nk}{\sqrt{p}} + \dfrac{10Nk\log p}{p})$ \\
\hline
TriOFM-$f_1$ & $\dfrac{nnz(\mathbf{A})4k + 12Nk^2 + 12Nk}{p}$ & $20\alpha \log p + \beta(\dfrac{8Nk}{\sqrt{p}} + \dfrac{10Nk\log p}{p})$ \\
\hline
OFM-$f_2$ & $\dfrac{nnz(\mathbf{A})4k + 16Nk^2 + 14Nk}{p}$ & $22\alpha \log p + \beta(\dfrac{8Nk}{\sqrt{p}} + \dfrac{12Nk\log p}{p})$ \\
\hline
TriOFM-$f_2$ & $\dfrac{nnz(\mathbf{A})4k + 16Nk^2 + 14Nk}{p}$ & $22\alpha \log p + \beta(\dfrac{8Nk}{\sqrt{p}} + \dfrac{12Nk\log p}{p})$ \\
\hline

\end{tabular}
}
\label{tb:complexity}
\end{table}

\begin{figure}
\begin{center}
\begin{tabular}{cc}
    \includegraphics[scale=0.20]{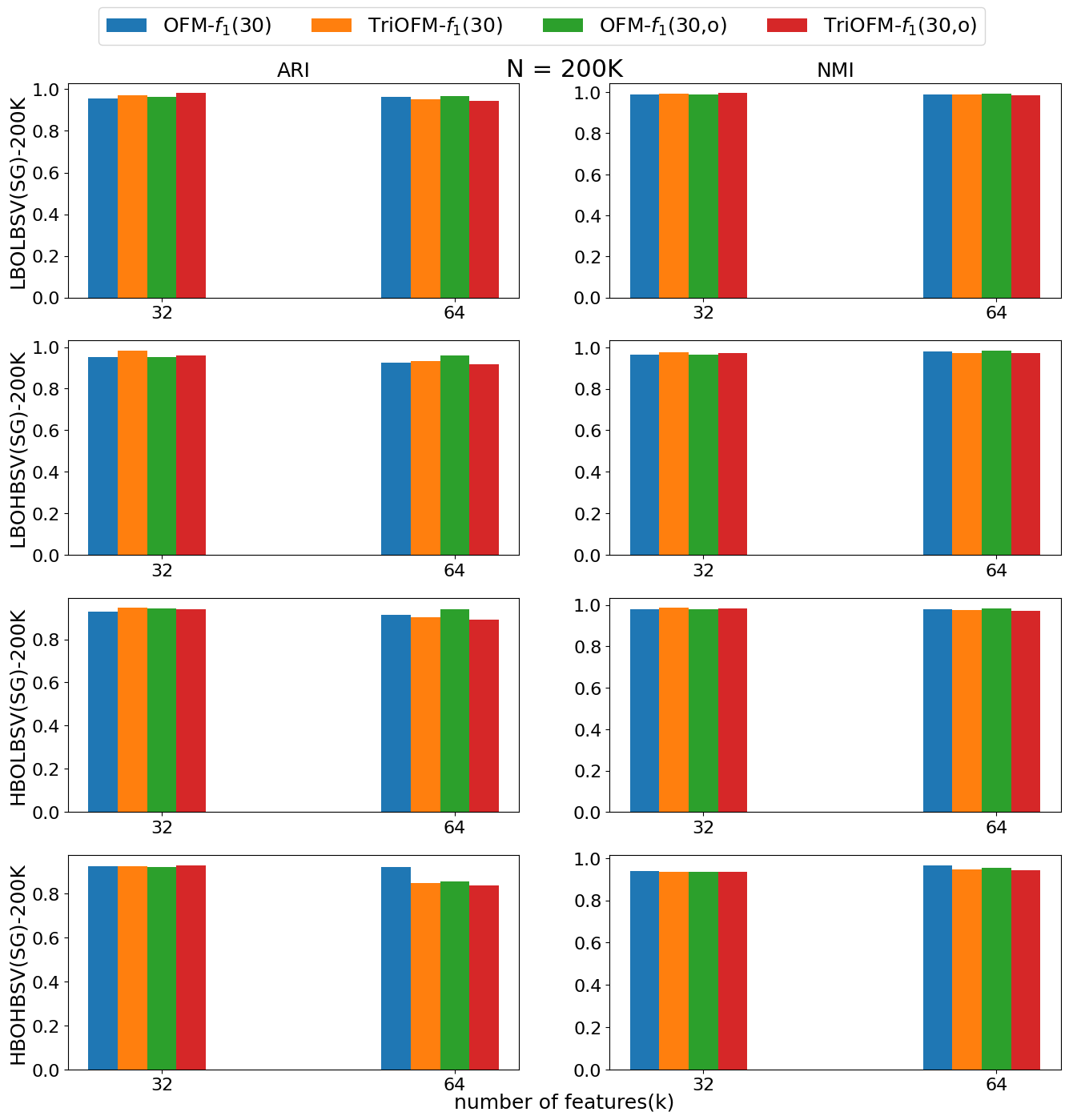} &
    \includegraphics[scale=0.20]{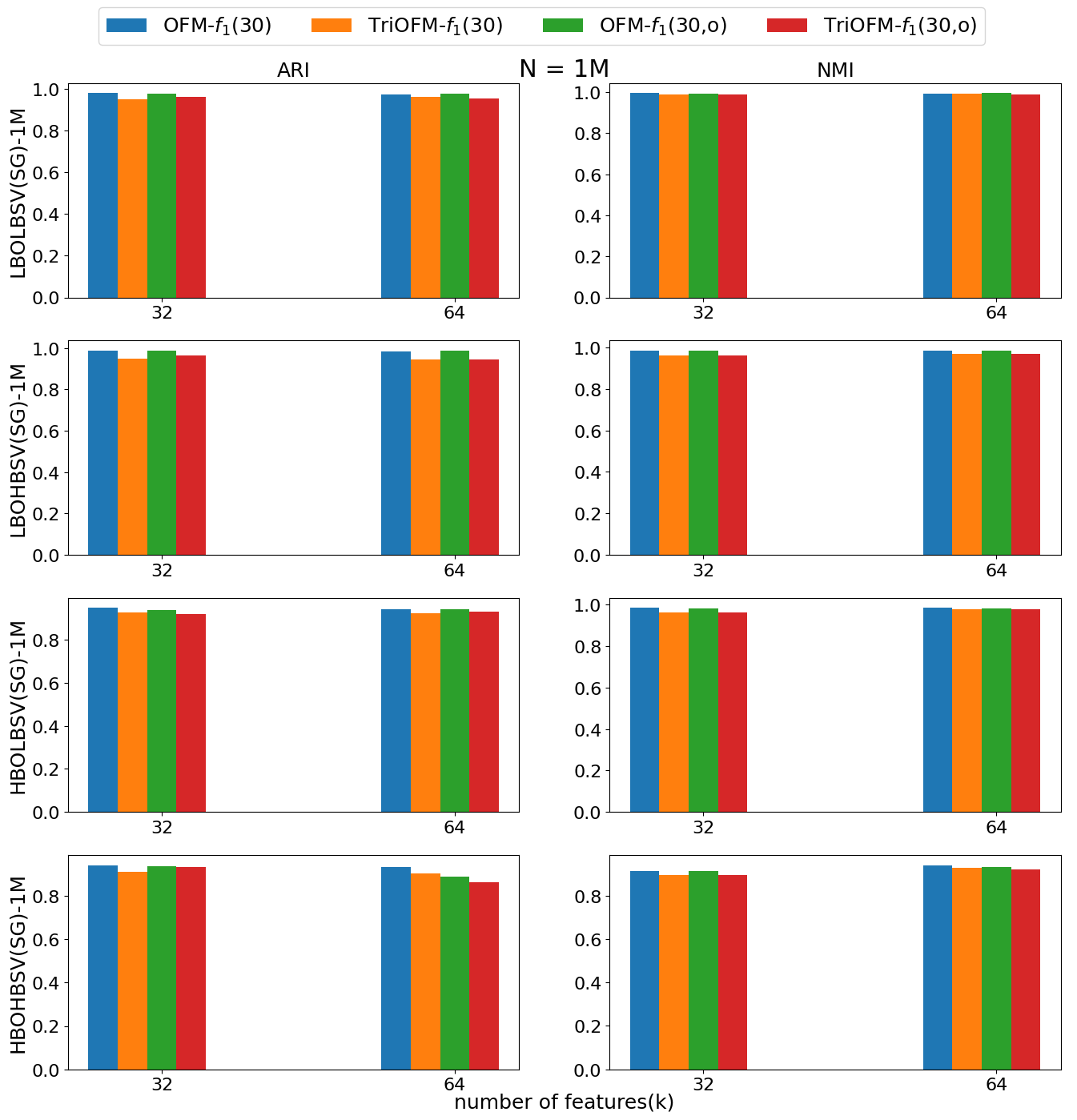}
\end{tabular}
\end{center}
\caption{Comparison between using the direct results of OFM-$f_1$ or TriOFM-$f_1$ and using the eigenvectors evaluated by OFM-$f_1$ or TriOFM-$f_1$ followed by orthogonalization and the Rayleigh-Ritz method, as features in spectral clustering quality. The static graphs used for comparisons have 200 thousand nodes (left) or 1 million nodes (right). `OFM-$f_1$(30)' indicates the features are the direct results of OFM-$f_1$ running with $30$ iterations. `OFM-$f_1$(30,o)' indicates the features are eigenvectors computed by `OFM-$f_1$(30)', orthogonalization, and the Rayleigh-Ritz method. Other notations in the plots have similar meanings.}
\label{fig:comporthf1}
\end{figure}

\begin{figure}
\begin{center}
\begin{tabular}{cc}
    \includegraphics[scale=0.20]{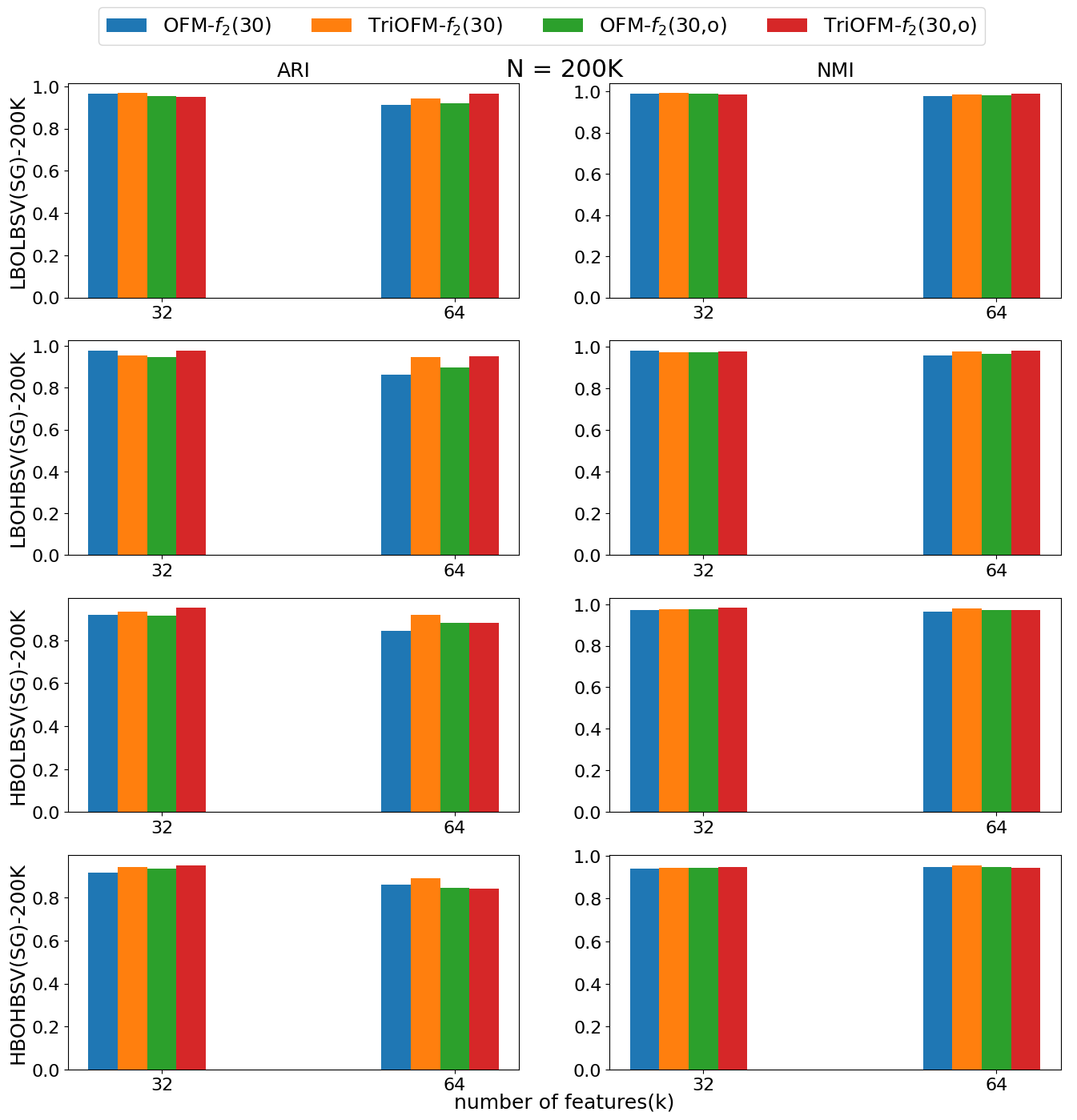} & 
    \includegraphics[scale=0.20]{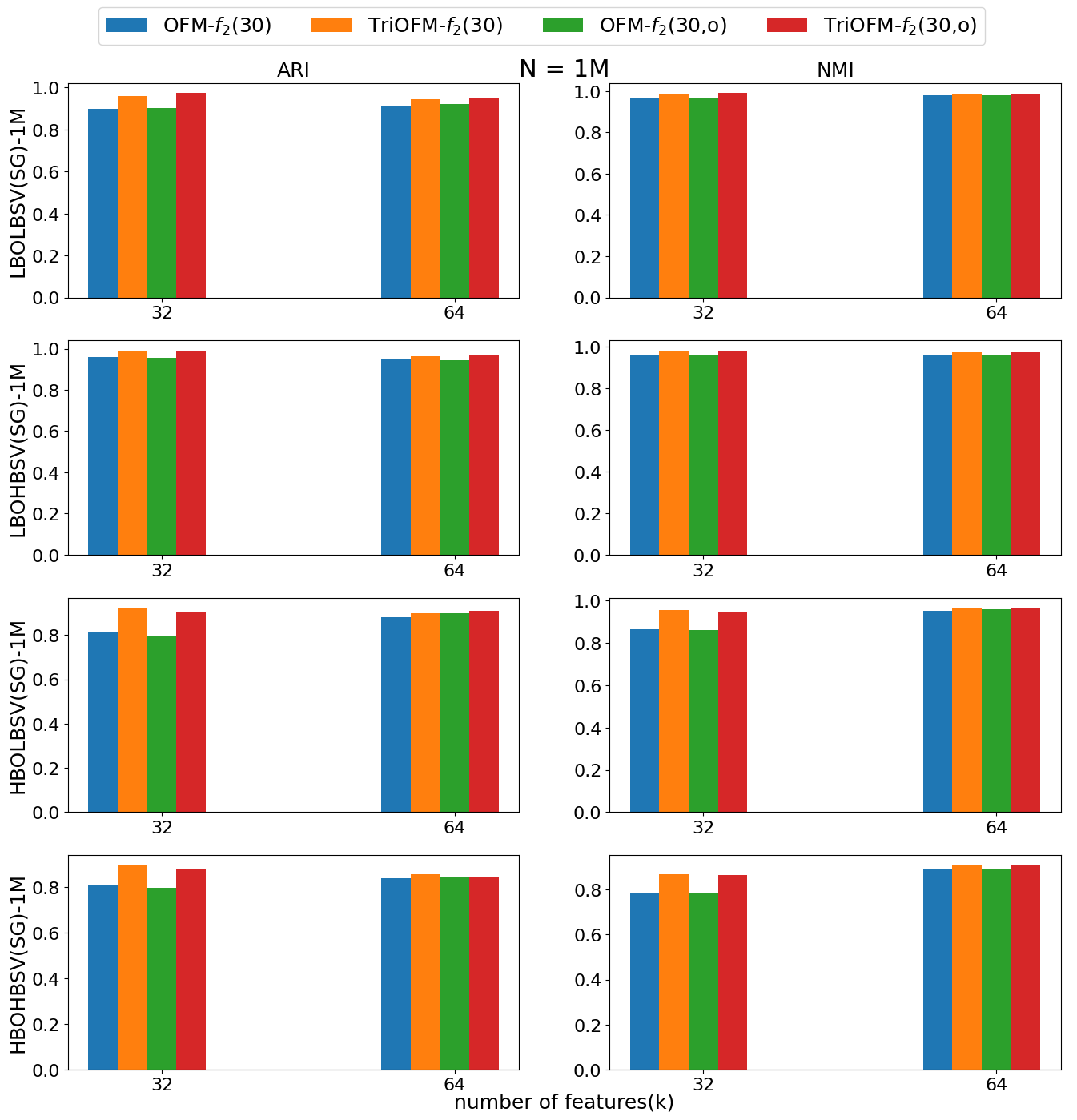}
\end{tabular}
\end{center}
\caption{Comparison between using the direct results of OFM-$f_2$ or TriOFM-$f_2$ and using the eigenvectors evaluated by OFM-$f_2$ or TriOFM-$f_2$ followed by orthogonalization and the Rayleigh-Ritz method, as features in spectral clustering quality. The static graphs used for comparisons have 200 thousand nodes (left) or 1 million nodes (right). `OFM-$f_2$(30)' indicates the features are the direct results of OFM-$f_2$ running with $30$ iterations. `OFM-$f_2$(30,o)' indicates the features are eigenvectors computed by `OFM-$f_2$(30)', orthogonalization, and the Rayleigh-Ritz method. Other notations in the plots have similar meanings.}
\label{fig:comporthf2}
\end{figure}

      

\begin{figure}[ht!]
  \begin{center}
    \begin{tabular}{c}
      \includegraphics[scale=0.34]{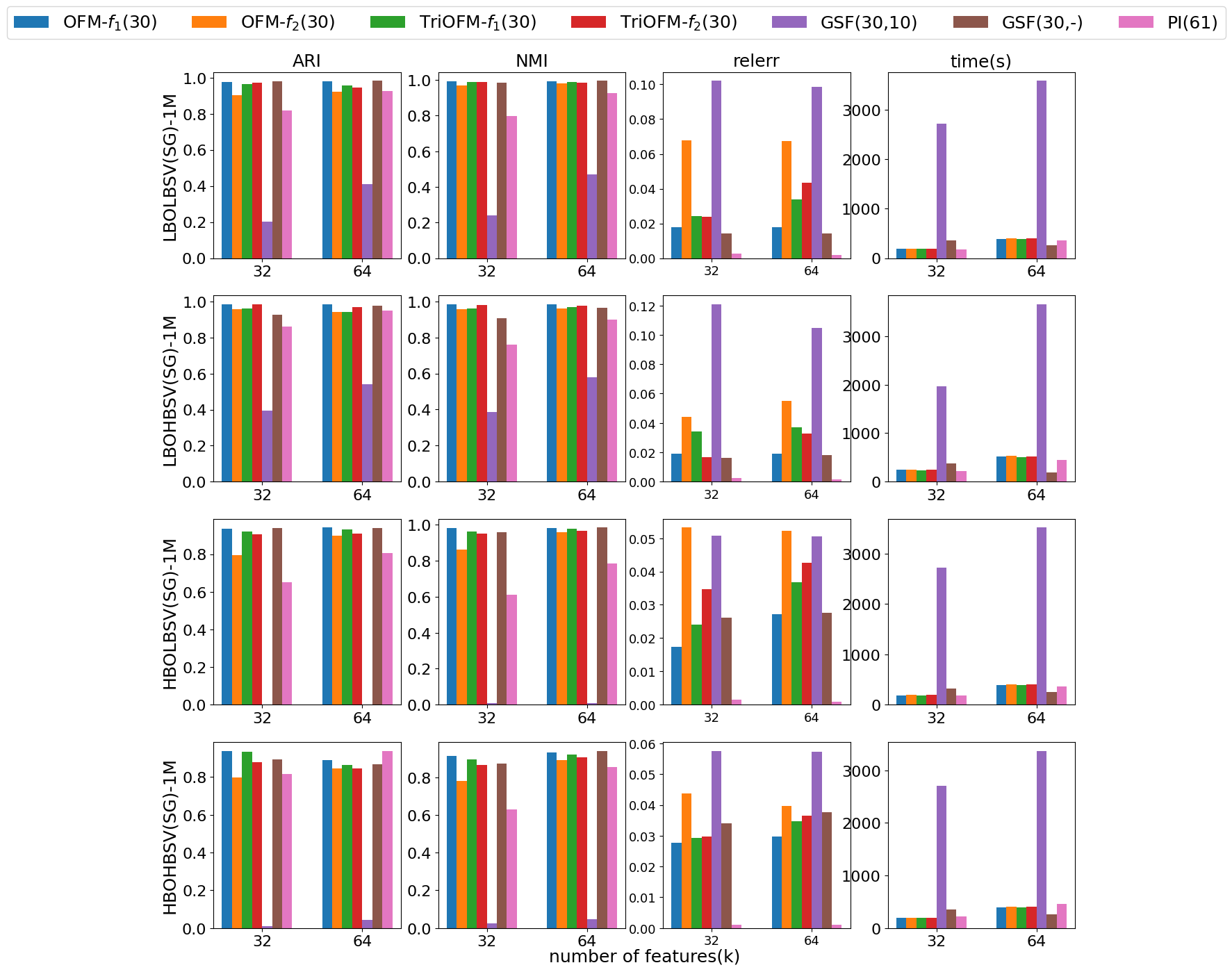} 
      
    \end{tabular}
  \end{center}
\caption{Comparison of the orthogonalization-free methods, GSF, and PI, on clustering static graphs each with $1$ million graph nodes. }
\label{fig:static-comp2}
\end{figure}

      

\begin{figure}[ht!]
  \begin{center}
    \begin{tabular}{c}
      \includegraphics[scale=0.34]{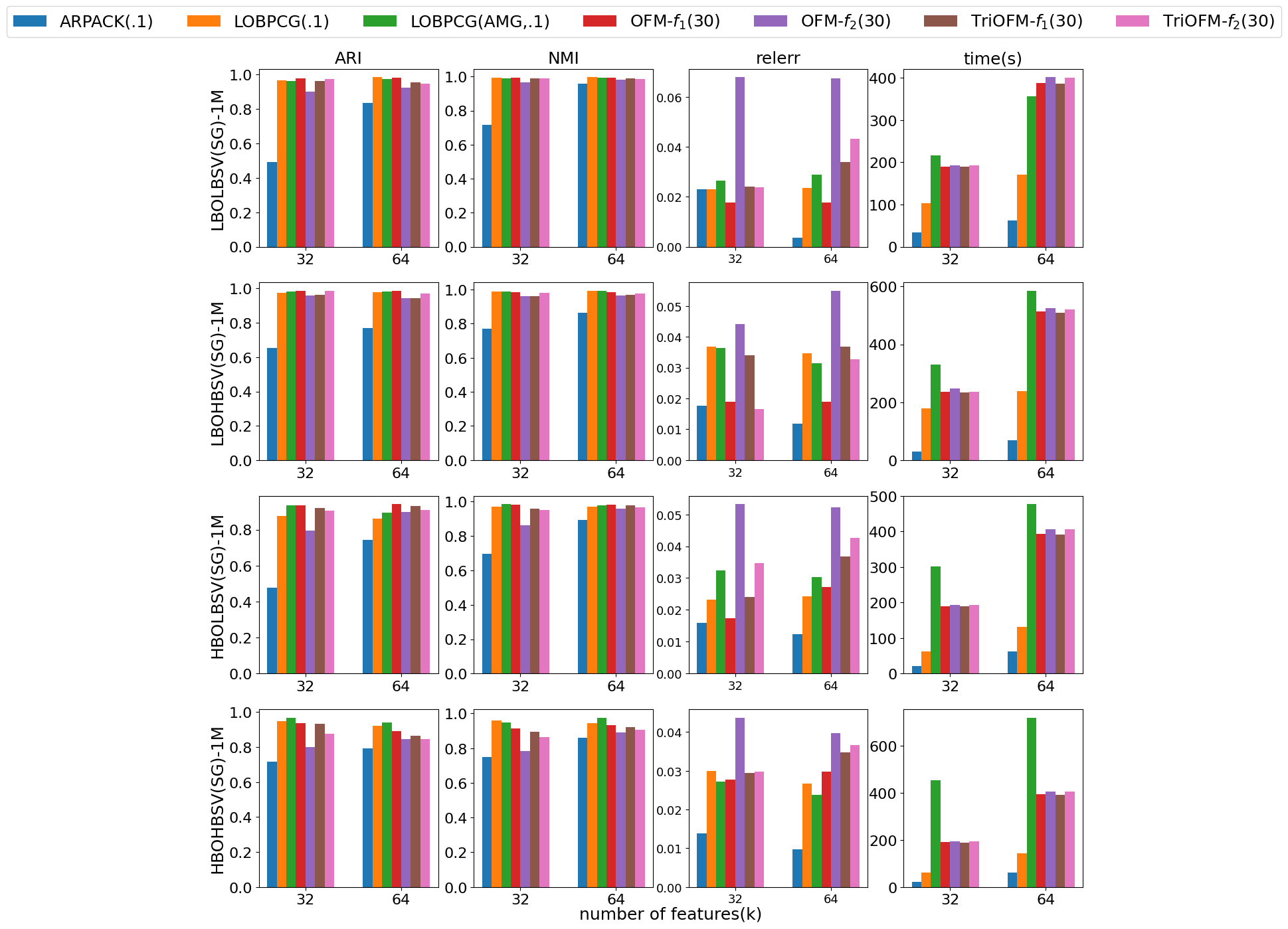} 
      
    \end{tabular}
  \end{center}
\caption{Comparison of the orthogonalization-free methods, ARPACK, LOBPCG with AMG preconditioning, and LOBPCG without preconditioning, on clustering static graphs each with $1$ million graph nodes. }
\label{fig:static-comp4}
\end{figure}

\section{Orthogonalization-Free Methods}
In this section, we describe four sequential orthogonalization-free methods. Implementation details will be described in the next section.

\subsection{OFM-$f_1$ and TriOFM-$f_1$}

We denote the eigenvalues of $\mathbf{A}$ as $-2 \leq \lambda_{1}\leq \lambda_{2} \leq ... \leq \lambda_{N} \leq 0$, and $\ell$ as the number of negative eigenvalues of a matrix, which will be used in later analysis. Let $\Lambda_\ell$ be the diagonal matrix with $\lambda_1,...,\lambda_\ell$ along the diagonal. Consider the $f_1$ optimization problem
\begin{equation}
    \min_{\mathbf{X}\in \mathbb{R}^{N\times k}} f_{1}(\mathbf{X}) = \|\mathbf{A}+\mathbf{X} \mathbf{X}^{T}\|_{F}^2,
\end{equation}
where the gradient 
\begin{equation}
\label{eq:gradientf1}
    \nabla f_{1}(\mathbf{X}) = 4 \mathbf{A}\mathbf{X} + 4\mathbf{X}\mathbf{X}^{T}\mathbf{X}.
\end{equation}
Local minima are computed in an iterative manner 
\begin{equation}
\label{eq:updatef1}
    \mathbf{X}^{(t+1)} = \mathbf{X}^{(t)} - \alpha \nabla f_{1}(\mathbf{X}^{(t)}),
\end{equation}
where $\alpha$ is the stepsize and superscript $t$ denotes the $t$-th iteration.

Note that $\mathbf{U}_{k}$ is not a local minimum of $f_{1}$ because $\nabla f_{1}(\mathbf{U}_k) \neq \mathbf{0}$. However, a local minimum $\mathbf{X}$ of $f_{1}$ is of a special form close to $\mathbf{U}_{k}$, which may be employed as a feature in spectral clustering. Objective $f_1$ has no spurious local minimum, and all local minima are global minima. The form of local minima has been proven in \cite{li2019coordinatewise, liu2015efficient} and summarized as follows in \cite{gao2022triangularized}: 

\begin{theorem}
    All stationary points of $f_{1}$ are of form $\mathbf{X} = \mathbf{U}_{\ell}\sqrt{-\Lambda_{\ell}}\mathbf{S}\mathbf{P}$ and all local minima are of form $\mathbf{X} = \mathbf{U}_{k}\sqrt{-\Lambda_{k}}\mathbf{Q}$, $\mathbf{S} \in \mathbf{R}^{\ell\times \ell}$ is a diagonal matrix with diagonal entries being $0$ or $1$ (at most $k$ $1$'s), $\mathbf{P} \in \mathbb{R}^{\ell \times k}$ and $\mathbf{Q} \in \mathbb{R}^{k \times k}$ are unitary matrices. Further, any local minimum is also a global minimum.
\end{theorem}

We use local minima of form $\mathbf{X} = \mathbf{U}_{k}\sqrt{-\Lambda_{k}}\mathbf{Q}$ as features in spectral clustering. Note that a unitary transformation $\mathbf{Q}$ preserves Euclidean distance, which indicates that using $\mathbf{X}$ is equivalent to using $\mathbf{U}_{k}\sqrt{-\Lambda_{k}}$ as a feature matrix for Euclidean distance based K-means clustering. Therefore, OFM-$f_1$ approximates the matrix $\mathbf{U}_{k}$ by a local minimum of $f_{1}$ which augments the eigenspace by weights $\sqrt{-\lambda_1}, ..., \sqrt{-\lambda_k}$. Though we do not have a theoretical analysis of if $\mathbf{U}_{k}\sqrt{-\Lambda_{k}}$ could replace $\mathbf{U}_{k}$ in spectral clustering, we observe effectiveness in numerical experiments. See Figure \ref{fig:compfeatures}. For details, please see the numerical section. Scaling the eigenvectors by the square root of the eigenvalues gives an embedding where $l_2$ distances in the embedding space approximate commute times between nodes on the graph \cite{qiu2007clustering, von2007tutorial}. We leave the theoretical analysis of the effectiveness of scaled eigenvectors and their possible relationship with commute times in spectral clustering to future works.

Another way to compute $\mathbf{X}$ is to recursively solve the single-column version of $f_1$. For example, we solve the single-column version of $f_{1}$ to get the local minimum $x_{1} = \sqrt{-\lambda_1}u_{1}$ where $\lambda_{1}$ and $u_{1}$ are the smallest eigenvalue and eigenvector, respectively. We apply the method to $\mathbf{A}_{2} = \mathbf{A} + x_{1}x_{1}^{T}$ to get $x_{2} = \sqrt{-\lambda_2}u_{2}$ where $\lambda_{2}$ and $u_{2}$ are the second smallest eigenvalue and eigenvector, respectively.
Note that we have $\nabla f_{1}(x_2) = 4(\mathbf{A}+x_{1}x_{1}^T)x_2 + 4 x_2 x_2^T x_2 = 4 \mathbf{A} x_2 + 4 (x_{1}x_{1}^T x_2 + x_2 x_2^T x_2)$. Similarly, while computing $x_{k} = \sqrt{-\lambda_k}u_{k}$ by applying the method to $\mathbf{A}_{k} = \mathbf{A} + \Sigma_{i=1}^{k-1} x_{i}x_{i}^{T}$, we have $\nabla f_{1}(x_k) = 4 \mathbf{A} x_k + 4 \Sigma_{i=1}^{k}x_i x_i^T x_k$. Based on this observation, Gao et al. \cite{gao2022triangularized} propose TriOFM-$f_1$, which adopts a new updating direction 
\begin{equation}
\label{eq:gradientg1}
    g_{1}(\mathbf{X}) = \mathbf{A}\mathbf{X} + \mathbf{X}triu(\mathbf{X}^{T}\mathbf{X})
\end{equation}
instead of the gradient $\nabla f_{1}(\mathbf{X})$. Their paper assumes that eigenvectors are sparse to apply TriOFM-$f_1$ as a coordinate-wise descent method. Though eigenvectors in spectral clustering are usually dense, we found that it is effective for spectral clustering. $g_{1}$ is not a gradient of any energy function; Gao et al. analyze the fixed point of 
\begin{equation}
\label{eq:updateg1}
    \mathbf{X}^{(t+1)} = \mathbf{X}^{(t)} - \alpha g_{1}(\mathbf{X}^{(t)}),
\end{equation}
in the following theorem \cite{gao2022triangularized}:

\begin{theorem}
    All fixed points of (\ref{eq:updateg1}) are of form $\mathbf{X} = \mathbf{U}_{\ell}\sqrt{-\Lambda_{\ell}}\mathbf{P}\mathbf{S}$ where $\mathbf{P} \in \mathbb{R}^{\ell \times k}$ is the first $k$ columns of an arbitrary $\ell\times \ell$ permutation matrix, and $\mathbf{S} \in \mathbf{R}^{k\times k}$ is a diagonal matrix with diagonal entries being $0$, $-1$, or $1$. Within these points all stable fixed points are of form $\mathbf{X} = \mathbf{U}_{k}\sqrt{-\Lambda_{k}}\mathbf{D}$, where $\mathbf{D} \in \mathbb{R}^{k\times k}$ is a diagonal matrix with diagonal entries being $1$ or $-1$. Others are unstable fixed points. 
\end{theorem}
Using a stable fixed point $\mathbf{X} = \mathbf{U}_{k}\sqrt{-\Lambda_{k}}\mathbf{D}$ is equivalent to using $\mathbf{U}_{k}\sqrt{-\Lambda_{k}}$ as a feature matrix for Euclidean distance based K-means clustering because the transformation $\mathbf{D}$ preserves Euclidean distance. Consequently, OFM-$f_1$ and TriOFM-$f_1$ essentially construct isomorphic features for K-means clustering. Again, we observe the effectiveness of using $\mathbf{U}_{k}\sqrt{-\Lambda_{k}}$ as features to replace $\mathbf{U}_k$ in spectral clustering in numerical results in Figure \ref{fig:compfeatures}. We hypothesize that using feature $\mathbf{U}_{k}\sqrt{-\Lambda_{k}}$ is effective in spectral clustering, and hence OFM-$f_1$ and TriOFM-$f_1$ are effective. Further tests and comparisons in the numerical section support this hypothesis.

\subsection{OFM-$f_2$ and TriOFM-$f_2$}

Now we consider the second optimization problem
\begin{equation}
    \min_{\mathbf{X}\in \mathbb{R}^{N\times k}} f_{2}(\mathbf{X}) =: tr((2\mathbf{I}-\mathbf{X}^{T}\mathbf{X})\mathbf{X}^{T}\mathbf{A}\mathbf{X})
\end{equation}
with gradient 
\begin{equation}
\label{eq:gradientf2}
    \nabla f_{2}(\mathbf{X}) = 4 \mathbf{A}\mathbf{X} - 2 \mathbf{X}\mathbf{X}^{T}\mathbf{A}\mathbf{X} - 2 \mathbf{A}\mathbf{X}\mathbf{X}^{T}\mathbf{X}.
\end{equation}
Local minima could be computed in an iterative manner
\begin{equation}
\label{eq:updatef2}
    \mathbf{X}^{(t+1)} = \mathbf{X}^{(t)} - \alpha \nabla f_{2}(\mathbf{X}^{(t)}),
\end{equation}
where $\alpha$ is the stepsize and the superscript $t$ denotes the $t$-th iteration.

Note that $\mathbf{U}_{k}$ is a local minimum because $\nabla f_{2}(\mathbf{U}_{k}) = \mathbf{0}$. OFM-$f_2$ directly evaluates iterations to locate any local minimum of $f_2$, which is a global minimum given the following theory proved by J. Lu and K. Thicke \cite{lu2017orbital}. 
\begin{theorem}
    All stationary points of OFM-$f_2$ are of form $\mathbf{X} = \mathbf{U}_{N}\mathbf{S}\mathbf{P}$ and all local minima are of form $\mathbf{X} = \mathbf{U}_{k}\mathbf{Q}$, where $S \in \mathbb{R}^{N\times N}$ is a diagonal matrix with diagonal entries being $0$ or $1$ (at most $k$ $1$'s), $\mathbf{P} \in \mathbb{R}^{N\times k}$ and $\mathbf{Q} \in \mathbb{R}^{k\times k}$ are unitary matrices. Further, any local minimum is also a global minimum.
\end{theorem}
Unlike OFM-$f_1$ and TriOFM-$f_1$, when used as the feature matrix to feed the Euclidean distance base K-means clustering, the local minimum $\mathbf{X} = \mathbf{U}_{k}\mathbf{Q}$ used in OFM-$f_2$ is equal to $\mathbf{U}_{k}$ only altered by a unitary transformation $\mathbf{Q}$ which preserves the Euclidean distance. 

Analogous to TriOFM-$f_1$, TriOFM-$f_2$ \cite{gao2022triangularized} employs
\begin{equation}
\label{eq:gradientg2}
    g_{2}(\mathbf{X}) = 2\mathbf{A}\mathbf{X} - \mathbf{A}\mathbf{X}triu(\mathbf{X}^{T}\mathbf{X}) - \mathbf{X}triu(\mathbf{X}^{T}\mathbf{A}\mathbf{X})
\end{equation}
as a new update rule for
\begin{equation}
\label{eq:updateg2}
    \mathbf{X}^{(t+1)} = \mathbf{X}^{(t)} - \alpha g_{2}(\mathbf{X}^{(t)}).
\end{equation}
The fixed points of (\ref{eq:updateg2}) are analyzed in the following theorem \cite{gao2022triangularized}:
\begin{theorem}
    All fixed points of (\ref{eq:updateg2}) are of form $\mathbf{X} = \mathbf{U}_{N}\mathbf{P}\mathbf{S}$ and all stable fixed points are of form $\mathbf{X} = \mathbf{U}_{k}\mathbf{D}$, where $\mathbf{P} \in \mathbb{R}^{\ell \times k}$ is the first $k$ columns of an arbitrary $N\times N$ permutation matrix, $\mathbf{S} \in \mathbf{R}^{k\times k}$ is a diagonal matrix with diagonal entries being $0$, $-1$, or $1$, and $\mathbf{D} \in \mathbf{R}^{k\times k}$ is a diagonal matrix with diagonal entries being $-1$, or $1$.
\end{theorem}
Therefore, the fix points constructed by TriOFM-$f_2$ is unitary to $\mathbf{U}_k$. 
In summary, compared to OFM-$f_1$ and TriOFM-$f_1$ whose convergent results are isomorphic to $\mathbf{U}_{k}\sqrt{-\Lambda_k}$, convergent results in OFM-$f_2$ and TriOFM-$f_2$ are directly isomorphic to $\mathbf{U}_{k}$, the desired eigenvectors in spectral clustering. The spectral clustering via orthogonalization-free methods has been summarized as Algorithm \ref{algo:spclustering}.

Note that none of the four optimization methods introduced above acquires orthogonalization, which is widely used in ARPACK, LOBPCG, block Chebyshev-Davidson method, PI-based method, and GSF. Figure \ref{fig:comporthf1} and \ref{fig:comporthf2} show that orthogonalizing the direct results from OFMs to get the eigenvectors does not improve clustering quality. Therefore, orthogonalization is unnecessary in OFMs, and OFMs are more scalable than eigensolvers like ARPACK and LOBPCG in parallel computing environments. See Figure \ref{fig:ofm-scalability-comp-cores} for an example.

In the numerical section, we will demonstrate the effectiveness of the orthogonalization-free methods for spectral clustering by comparing them with other methods, especially the advantages in the streaming graph scenario and scalability in parallel computing environments. Before proceeding with the numerical results, we describe the implementation details and complexity analysis.

      


\section{Implementation Details}
In this section, we describe how we implement the orthogonalization-free methods, including momentum acceleration, conjugate gradient for momentum parameters, and line search for optimal stepsizes. We will also delineate parallel implementation details of each critical component of the methods, especially sparse matrix times tall-skinny matrix multiplication. 

\subsection{Momentum Acceleration}
In traditional gradient descent methods, momentum is widely used to smooth oscillatory trajectories and accelerate convergence \cite{qian1999momentum}. Instead of moving along the gradient direction directly, the momentum method moves along an accumulation of gradient directions with a discounting parameter $\beta \in (0,1]$, i.e., 
\begin{equation}
    \mathbf{V}^{(t)} = \beta g(\mathbf{X}^{(t)}) + (1 - \beta) \mathbf{V}^{(t-1)},
\end{equation}
where $\mathbf{V}^{t}$ denotes the accumulation direction and $g$ is one of $\{\nabla f_{1}, g_{1}, \nabla f_{2}, g_{2}\}$. Then the iteration moves along $\mathbf{V}^{(t)}$ to update $\mathbf{X}^{(t+1)}$ with stepsize $\alpha$, i.e., $\mathbf{X}^{(t+1)} = \mathbf{X}^{(t)} + \alpha \mathbf{V}^{(t)}$. To avoid manually choosing the momentum parameter $\beta$, we adopt the idea of conjugate gradient (CG) \cite{golub2013matrix}, a momentum method with adaptive momentum parameters. F. Corsetti \cite{corsetti2014orbital} adopts a nonlinear CG in optimizing OFM-$f_2$, and Gao et al. \cite{gao2022triangularized} adopts a column-wise nonlinear CG, which is an extension to Polak-Reeves CG \cite{polak1969note}, in TriOFM-$f_{1}$ and TriOMM-$f_{2}$. We adopt the same columnwise nonlinear CG in OFM-$f_{1}$, TriOFM-$f_{1}$, OFM-$f_{2}$, and TriOFM-$f_{2}$.



\subsection{Stepsizes}
In previous sections, we describe algorithms with a constant stepsize for simplicity. However, choosing a stepsize $\alpha$ is usually tedious, and a terrible choice of $\alpha$ probably impairs the convergence. Thus, we adopt an exact line search strategy to determine the optimal stepsizes automatically. We can conduct the exact line search by minimizing quartic polynomials $f_{1}$ and $f_{2}$, or equivalently, calculating the stable points of the derivative cubic polynomials. Taking OFM-$f_{1}$ as an example, the cubic polynomial is
    \begin{align}
    \label{eq:f1-derivative}
        \frac{d}{d\alpha} f_{1}(\mathbf{X}+\alpha \mathbf{V}) &= tr(\mathbf{V}^T \nabla f_{1}(\mathbf{X}+\alpha \mathbf{V})) \nonumber\\
        &= \alpha^3 tr((\mathbf{V}^T \mathbf{V})^2) + 3 \alpha^2 tr(\mathbf{V}^T \mathbf{V} \mathbf{X}^T \mathbf{V}) \nonumber\\
        &+ \alpha tr(\mathbf{V}^T \mathbf{A} \mathbf{V} + (\mathbf{V}^T \mathbf{X})^2 + \mathbf{V}^T \mathbf{X} \mathbf{X}^T \mathbf{V} + \mathbf{V}^T \mathbf{V} \mathbf{X}^T \mathbf{X}) \nonumber\\
        &+ tr(\mathbf{V}^T \mathbf{A} \mathbf{X}) + tr(\mathbf{V}^T \mathbf{X}\mathbf{X}^T \mathbf{X}).
    \end{align}

Solving the above equation might result in one, two, or three real roots. There are three different choices of $\alpha$:
\begin{itemize}
    \item the real root if only one real root exists;
    \item the real root of single multiplicity if two real roots exist;
    \item the real root which achieves the minimal function value when three real roots exist.
\end{itemize}
We could efficiently evaluate a cubic polynomial's real roots via Cardano's formula. Though the evaluated stepsize does not work for TriOFM-$f_{1}$, Gao et al. \cite{gao2022triangularized} show that we could compute the stepsize for TriOFM-$f_{1}$ in a similar manner by solving $tr(\mathbf{V}_{i}^T g_{1}(\mathbf{X}_{i}+\alpha_i \mathbf{V}_{i})) = 0, \forall i$ instead, which can be expressed as again a cubic polynomial of $\alpha_i$:

    \begin{align}
    \label{eq:alphai-f1}
        tr(\mathbf{V}_{i}^T g_{1}(\mathbf{X}_{i}+\alpha_i \mathbf{V}_{i})) &= \alpha_i^3 tr(\mathbf{V}_{i}^T \mathbf{V}_{i} triu(\mathbf{V}_{i}^T \mathbf{V}_{i})) \nonumber\\
        &+ \alpha_i^2 tr(\mathbf{V}_{i}^T \mathbf{V}_{i} triu(\mathbf{X}_{i}^T \mathbf{V}_{i}) + \mathbf{V}_{i}^T \mathbf{V}_{i} triu(\mathbf{V}_{i}^T \mathbf{X}_{i}) + \mathbf{V}_{i}^T \mathbf{X}_{i} triu(\mathbf{V}_{i}^T \mathbf{V}_{i})) \nonumber\\
        &+ \alpha_i tr(\mathbf{V}_{i}^T \mathbf{A} \mathbf{V}_{i} + \mathbf{V}_{i}^T \mathbf{X}_{i} triu(\mathbf{V}_{i}^T \mathbf{X}_{i}) + \mathbf{V}_{i}^T \mathbf{X}_{i} triu(\mathbf{X}_{i}^T \mathbf{V}_{i}) 
        + \mathbf{V}_{i}^T \mathbf{V}_{i} triu(\mathbf{X}_{i}^T \mathbf{X}_{i})) \nonumber\\
        &+ tr(\mathbf{V}_{i}^T \mathbf{A} \mathbf{X}_{i} + \mathbf{V}_{i}^T \mathbf{X}_{i} triu(\mathbf{X}_{i}^T \mathbf{X}_{i})).
    \end{align}

Analogous to OFM-$f_1$, we compute the stepsizes in OFM-$f_2$ by evaluating the roots of
    \begin{align}
    \label{eq:f2-derivative}
        \frac{d}{d\alpha} f_{2}(\mathbf{X}+\alpha \mathbf{V}) &= tr(\mathbf{V}^T \nabla f_{2}(\mathbf{X}+\alpha \mathbf{V}))\nonumber \\
        &= -4\alpha^3 tr(\mathbf{V}^T \mathbf{V} \mathbf{V}^T \mathbf{A} \mathbf{V}) - 6 \alpha^2 tr(\mathbf{V}^T \mathbf{X} \mathbf{V}^T \mathbf{A} \mathbf{V} + \mathbf{V}^T \mathbf{V} \mathbf{V}^T \mathbf{A} \mathbf{X})\nonumber \\
        &+ 2\alpha tr(2\mathbf{V}^T \mathbf{A} \mathbf{V} - 2\mathbf{V}^T \mathbf{X} \mathbf{X}^T \mathbf{A} \mathbf{V} - 2\mathbf{V}^T \mathbf{X} \mathbf{V}^T \mathbf{A} \mathbf{X} - \mathbf{V}^T \mathbf{V} \mathbf{X}^T \mathbf{A} \mathbf{X} - \mathbf{X}^T \mathbf{X} \mathbf{V}^T \mathbf{A} \mathbf{V}) \nonumber\\
        &+ 2 tr(2\mathbf{V}^T \mathbf{A} \mathbf{X} - \mathbf{V}^T \mathbf{X} \mathbf{X}^T \mathbf{A} \mathbf{X} - \mathbf{X}^T \mathbf{X} \mathbf{V}^T \mathbf{A} \mathbf{X}).
    \end{align}
Like TriOFM-$f_1$, TriOFM-$f_2$ evaluates the stepsizes by solving $tr(\mathbf{V}_{i}^T g_{2}(\mathbf{X}_{i}+\alpha_i \mathbf{V}_{i})) = 0, \forall i$, where
    \begin{align}  
    \label{eq:alphai-f2}
        tr(\mathbf{V}_{i}^T g_{2}(\mathbf{X}_{i}+\alpha_i \mathbf{V}_{i})) &= -\alpha_i^3 tr(\mathbf{V}_{i}^T \mathbf{A} \mathbf{V}_{i} triu(\mathbf{V}_{i}^T \mathbf{V}_{i}) + \mathbf{V}_{i}^T \mathbf{V}_{i} triu(\mathbf{V}_{i}^T \mathbf{A} \mathbf{V}_{i})) \nonumber\\
        &- \alpha_i^2 tr(\mathbf{V}_{i}^T \mathbf{A} \mathbf{X}_{i} triu(\mathbf{V}_{i}^T \mathbf{V}_{i}) + \mathbf{V}_{i}^T \mathbf{A} \mathbf{V}_{i} triu(\mathbf{X}_{i}^T \mathbf{V}_{i}) + \mathbf{V}_{i}^T \mathbf{A} \mathbf{V}_{i} triu(\mathbf{V}_{i}^T \mathbf{X}_{i})\nonumber \\
        &+ \mathbf{V}_{i}^T \mathbf{X}_{i} triu(\mathbf{V}_{i}^T \mathbf{A} \mathbf{V}_{i}) + \mathbf{V}_{i}^T \mathbf{V}_{i} triu(\mathbf{X}_{i}^T \mathbf{A} \mathbf{V}_{i}) + \mathbf{V}_{i}^T \mathbf{V}_{i} triu(\mathbf{V}_{i}^T \mathbf{A} \mathbf{X}_{i}))\nonumber \\
        &+ \alpha_i tr(2\mathbf{V}_{i}^T \mathbf{A} \mathbf{V}_{i} - \mathbf{V}_{i}^T \mathbf{A} \mathbf{X}_{i} triu(\mathbf{X}_{i}^T \mathbf{V}_{i}) - \mathbf{V}_{i}^T \mathbf{A} \mathbf{X}_{i} triu(\mathbf{V}_{i}^T \mathbf{X}_{i}) \nonumber\\
        &- \mathbf{V}_{i}^T \mathbf{A} \mathbf{V}_{i} triu(\mathbf{X}_{i}^T \mathbf{X}_{i}) - \mathbf{V}_{i}^T \mathbf{X}_{i} triu(\mathbf{X}_{i}^T \mathbf{A} \mathbf{V}_{i}) \nonumber\\
        &- \mathbf{V}_{i}^T \mathbf{X}_{i} triu(\mathbf{V}_{i}^T \mathbf{A} \mathbf{X}_{i}) - \mathbf{V}_{i}^T \mathbf{V}_{i} triu(\mathbf{X}_{i}^T \mathbf{A} \mathbf{X}_{i})) \nonumber\\
        &+ tr(2\mathbf{V}_{i}^T \mathbf{A} \mathbf{X}_{i} - \mathbf{V}_{i}^T \mathbf{A} \mathbf{X}_{i} triu(\mathbf{X}_{i}^T \mathbf{X}_{i}) - \mathbf{V}_{i}^T \mathbf{X}_{i} triu(\mathbf{X}_{i}^T \mathbf{A} \mathbf{X}_{i})).
    \end{align}

By combining conjugate gradient and exact line search, we summarize the entire framework of the orthogonalization-free method as Algorithm \ref{algo:framework}, where $\odot$ denotes element-wise multiplication, $\oslash$ denotes element-wise division, and $sumeachcol(\mathbf{G})$ sums each column of $\mathbf{G}$ and returns a column vector. 

\begin{algorithm}
\caption{Orthogonalization-free algorithm framework}
\label{algo:framework}
\begin{algorithmic}[1]
\State Input: symmetric matrix $\mathbf{A}$, initial point $\mathbf{X}^{(0)}$, stepsize $\alpha$, maximum iteration number $t_{max}$
\State Output: $\mathbf{X}^{(t_{max})}$
\State $\mathbf{G}^{(0)} = g(\mathbf{X}^{(0)})$
\State $\mathbf{V}^{(0)} = -\mathbf{G}^{(0)}$
\State $\mathbf{X}^{(1)} = \mathbf{X}^{(0)} + \alpha \mathbf{V}^{(0)}$
\State $t = 1$
\While{ $t < t_{max}$}
\State $\mathbf{G}^{(t)} = g(\mathbf{X}^{(t)})$
\State $\mathbf{\beta}^{(t)} = sumeachcol((\mathbf{G}^{(t)}-\mathbf{G}^{(t-1)})\odot \mathbf{G}^{(t)}) \oslash sumeachcol(\mathbf{G}^{(t-1)}\odot \mathbf{G}^{(t-1)})$
\State Compute $\mathbf{V}_{i}^{(t)} = -\mathbf{G}_{i}^{(t)} + \mathbf{\beta}^{(t)}_{i}\mathbf{V}_{i}^{(t-1)}$, $\forall i$
\State Evaluate stepsizes $\alpha^{(t)}$ by exact line search
\State Update $\mathbf{X}_{i}^{(t+1)} = \mathbf{X}_{i}^{(t)} + \mathbf{\alpha}^{(t)}_{i}\mathbf{V}_{i}^{(t)}$, for $\forall i$
\State $t = t+1$
\EndWhile
\end{algorithmic}

\end{algorithm}

\begin{figure}[ht!]
  \begin{center}
    \begin{tabular}{c}
      \includegraphics[scale=0.34]{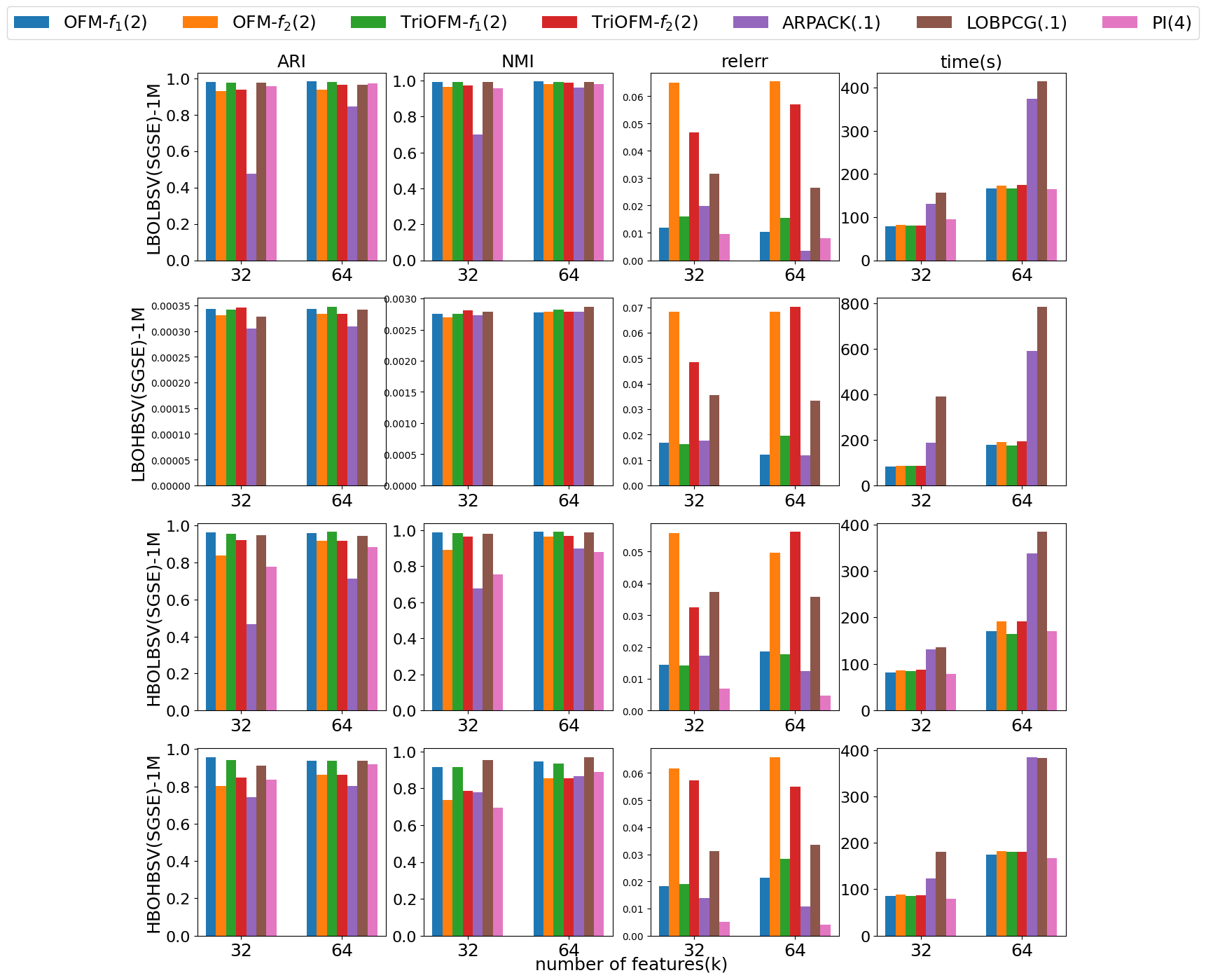} 
      
    \end{tabular}
  \end{center}
\caption{Comparison of the orthogonalization-free methods, PI, ARPACK, and LOBPCG, on clustering edge-sampling streaming graphs each with $1$ million graph nodes. }
\label{fig:streaming-comp-1M-e}
\end{figure}

\begin{figure}[ht!]
  \begin{center}
    \begin{tabular}{c}
      \includegraphics[scale=0.34]{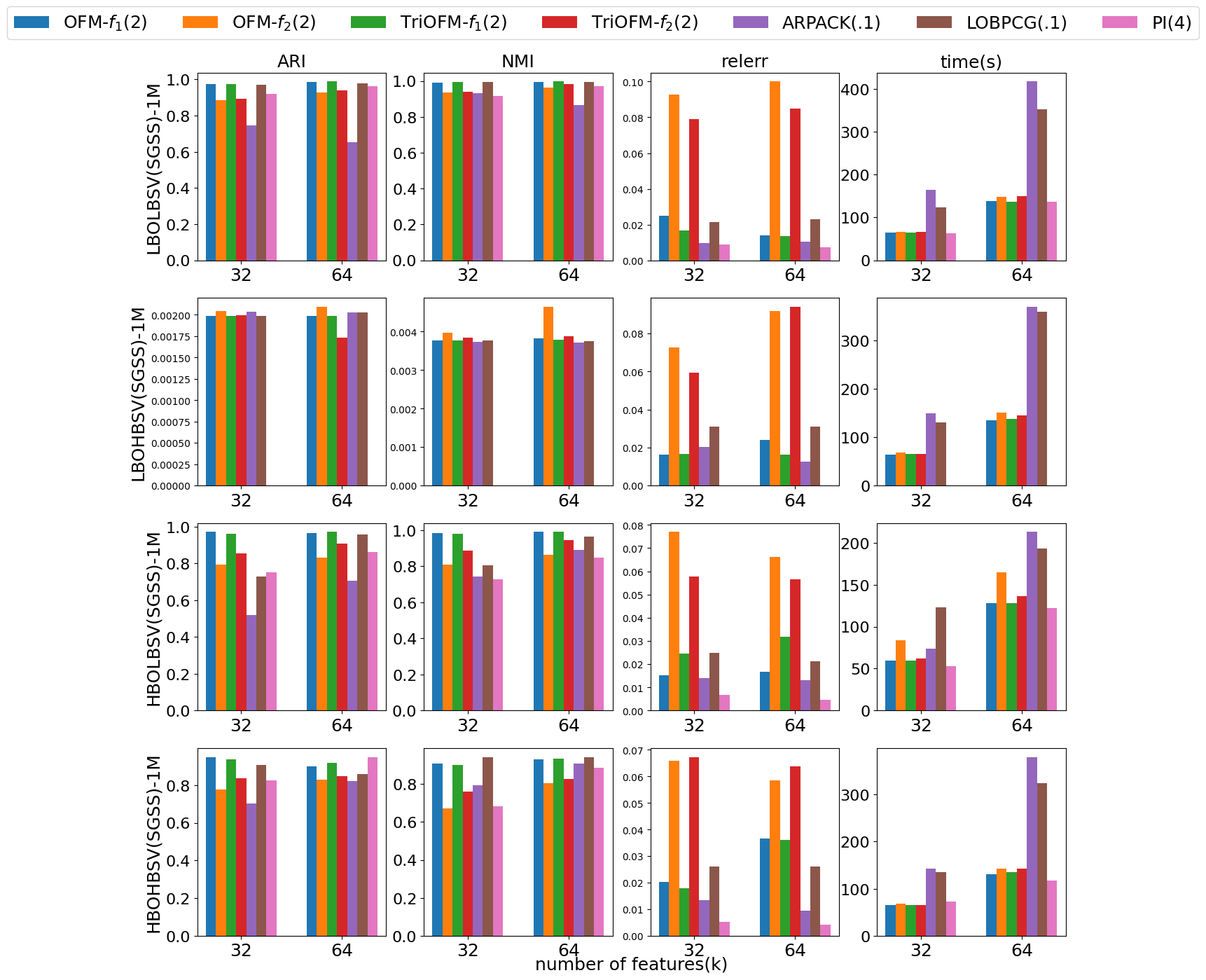}
      
    \end{tabular}
  \end{center}
\caption{Comparison of the orthogonalization-free methods, PI, ARPACK, and LOBPCG, on clustering snowball-sampling streaming graphs each with $1$ million graph nodes. }
\label{fig:streaming-comp-1M-s}
\end{figure}

\begin{figure}
\centering
\begin{tikzpicture}
  \node at (0.5,-0.5) {$\mathbf{Y}$};
  \node at (3.5,-0.5) {$\mathbf{A}$};
  \node at (6.5,-0.5) {$\mathbf{X}$};
  \node at (1.5,1.5) {=};
  \node at (5.5,1.5) {$\times$};
  \draw[help lines] (0,0) grid (1, 3);
  \draw (0,0.33) -- (1,0.33);
  \draw (0,0.67) -- (1,0.67);
  \draw (0,1.33) -- (1,1.33);
  \draw (0,1.67) -- (1,1.67);
  \draw (0,2.33) -- (1,2.33);
  \draw (0,2.67) -- (1,2.67);
  \node at (0.5, 0.5) {$\mathbf{Y}[7]$};
  
  \draw[help lines] (2,0) grid (5,3);
  \node at (3.5,0.5) {$\mathbf{A}[2,1]$};
  
  \draw[help lines] (6,0) grid (7, 3);
  \draw (6,0.33) -- (7,0.33);
  \draw (6,0.67) -- (7,0.67);
  \draw (6,1.33) -- (7,1.33);
  \draw (6,1.67) -- (7,1.67);
  \draw (6,2.33) -- (7,2.33);
  \draw (6,2.67) -- (7,2.67);
  \node at (6.5, 1.166) {$\mathbf{X}[5]$};
\end{tikzpicture}
\caption{We borrow the Figure 1 in \cite{bchdav4clustering2022} as an illustration of A-Stationary 1.5D SpMM $\mathbf{Y} = \mathbf{A}\mathbf{X}$ when the number of processes is $p = 9$. The $P(2,1)$ process owns the blocks $\mathbf{Y}[7], \mathbf{X}[5]$, and $\mathbf{A}[2,1]$.} 
\label{fig:spmm}
\end{figure}

\subsection{Parallelization}
To handle large-scale graphs, we parallelize and accelerate computation using multiple processes and threads, i.e., hybrid implementation. Multithreading implementation for a single node is straightforward by calling BLAS libraries. In our case, we directly call OpenBLAS library \cite{xianyi2012model} to accelerate dense-dense matrix computation and Intel MKL library \cite{wang2014intel} for sparse-dense matrix multiplication. Multiprocessing implementation for data distributed among multiple compute nodes requires cautious designs and tremendous efforts, which is the main focus of this section. 

The first critical component we need to parallelize is performing distributed sparse matrix times tall-and-skinny matrix (SpMM) computation in the form of
\begin{equation}
    \label{eq:spmm}
    \mathbf{Y} = \mathbf{A} \mathbf{X},
\end{equation}
where $\mathbf{X}$ and $\mathbf{Y}$ are of size $N \times k$, and $k \ll N$. Here, matrices $\mathbf{A}, \mathbf{X}$ and $\mathbf{Y}$ are distributed among multiple compute nodes. SpMM is fundamental in a variety of computation and engineering problems, for instance \cite{pang2020interpolative, bremer2021fast, tu2022linear}.
To perform distributed SpMM, we employ the A-Stationary 1.5D algorithm in \cite{selvitopi2021distributed}. 
Assume a 2D process/node grid $P$ of size $\sqrt{p} \times \sqrt{p}$. The algorithm partitions the sparse matrix $\mathbf{A}$ into a 2D block grid of size $\sqrt{p} \times \sqrt{p}$ and the dense matrices $\mathbf{X}$ and $\mathbf{Y}$ into $p$ row blocks, such that the $(i,j)$-th process $P(i,j)$ owns the $(i,j)$-th block $\mathbf{A}[i,j]$ of $\mathbf{A}$, the $(j\sqrt{p} + i)$-th block $\mathbf{X}[j\sqrt{p} + i]$ of $\mathbf{X}$, and the $(i\sqrt{p} + j)$-th block $\mathbf{Y}[i\sqrt{p} + j]$ of $\mathbf{Y}$. See Figure \ref{fig:spmm} for an illustration. The algorithm first replicates $\mathbf{X}$ among $\sqrt{p}$ processes in each column of the process grid, such that $P(i, j)$ has $\sqrt{p}$ blocks of $\mathbf{X}$, which are given by $\mathbf{X}[j\sqrt{p} + \ell]$ for $0 \leq \ell < \sqrt{p}$. Next, the processes perform the local computation of form $\mathbf{Y}^{j}[i\sqrt{p}+\ell] = \mathbf{A}[i,j]\mathbf{X}[j\sqrt{p}+\ell]$ for $0 \leq \ell < \sqrt{p}$ with $\mathbf{Y}^{j}$ denoting the partial dense resulting matrix evaluated by the process at $j$-th column of the grid. In the end, each process sums up the partial dense matrices to get the final result matrix at each process with $\mathbf{Y}[i\sqrt{p} + j] = \Sigma_{\ell=0}^{\sqrt{p}} \mathbf{Y}^{\ell}[i\sqrt{p} + j]$. For more details, please refer to \cite{selvitopi2021distributed}. On average, each process requires $nnz(\mathbf{A})/p + 2Nk/\sqrt{p}$ cost to store data, where $nnz(\mathbf{A})$ is the number of nonzeros in $\mathbf{A}$, and $nnz(\mathbf{A})2k/p$ flops to perform local computations.

To analyze communication cost, we assume that it costs $\alpha + \beta w$ time, where $\alpha$ is the latency and $\beta$ is the reciprocal bandwidth, to send a message of size $w$ from one process/node to another \cite{selvitopi2021distributed}.
The MPI\_Allgather collective is applied to realize the replication among $\sqrt{p}$ processes in each column of the process grid. If implemented with a recursive doubling algorithm, the MPI\_Allgather collective at each process takes $\alpha \log p + \beta w p$ cost to gather $w$ words from all processes in the same communicator  \cite{chan2007collective}. Before the replication for each grid column in the A-Stationary 1.5D algorithm, each process has $Nk/p$ words, and thus the respective MPI\_Allgather has a cost of $\alpha \log p + \beta Nk/\sqrt{p}$. The MPI\_Reduce\_scatter collective with a summation operator is adopted to realize the reduction of partial dense matrices among $\sqrt{p}$ processes in a row of the 2D process grid. With a recursive halving implementation, the MPI\_Reduce\_scatter collective takes $\alpha \log p + \beta w$ cost to collect $w$ words from all process in the same communicator and then scatter $w/p$ words back to each process \cite{chan2007collective}. In our case, the respective MPI\_Reduce\_scatter costs $\alpha \log p + \beta Nk/\sqrt{p}$ to reduce and scatter $Nk/\sqrt{p}$ words at each process. Therefore, the total communication cost for an SpMM is 
\begin{equation}
\label{eq:communicationcost-spmm}
2\alpha \log p + \beta \dfrac{2Nk}{\sqrt{p}}.
\end{equation}
One may notice that, after evaluating a distributed SpMM, $\mathbf{X}$ and $\mathbf{Y}$ are distributed differently, which induces barriers in further applying $\mathbf{A}$ to $\mathbf{Y}$. To handle this, we transform the process grid first and then implicitly apply the identity matrix to $\mathbf{Y}$ so that $\mathbf{Y}$ and $\mathbf{X}$ share the same distribution. This additional step does not acquire local computation but doubles the communication cost.

Besides SpMM, our algorithms involve many dot-products in the form of $\mathbf{Y}^T \mathbf{X}$. Given the partitions of $\mathbf{Y}$ and $\mathbf{X}$, it is natural and efficient first to evaluate the dot-products of the local portions of $\mathbf{X}$ and $\mathbf{Y}$ at each process/node, then sum up all the local products and broadcast the results to every process involved. The local computation takes $2N k^2 / p$ flops. We use MPI\_Allreduce to realize the summation of local products and broadcast the results. If implemented using a tree algorithm, the MPI\_Allreduce collective reduces $w$ words from all processes at a single process and broadcasts the result to all processes in the same communicator, which has $2\alpha \log p + 2\beta w \log p$ \cite{chan2007collective}. Each process has $Nk/p$ words during the computation of $\mathbf{Y}^T \mathbf{X}$, thus the communication cost the MPI\_Allreduce collective takes is 
\begin{equation}
\label{eq:communicationcost-dot}
2\alpha \log p + 2\beta \dfrac{Nk}{p} \log p.
\end{equation}

\section{Complexity Analysis}
This section analyzes each algorithm's per-iteration time complexity, assuming multi-processing and single-threading implementation. We only analyze the time complexity of the dominating steps $7-11$ in Algorithm \ref{algo:framework}.

\subsection*{Complexity Analysis of Step 7}
The computation $\mathbf{G}^{(t)} = g(\mathbf{X}^{(t)})$ involves an SpMM, multiple dot-products, and rescaling if necessary. Note that $\mathbf{X}^T \mathbf{A} \mathbf{X}$ is computed by first performing a SpMM $\mathbf{A} \mathbf{X}$ and then dot-products $\mathbf{X}^T (\mathbf{A} \mathbf{X})$. And $\mathbf{X} \mathbf{X}^T \mathbf{X}$ is computed by first performing $\mathbf{X}^T \mathbf{X}$ and then purely local computation $\mathbf{X} (\mathbf{X}^T \mathbf{X})$. Other analogous computations are carried out in the same way as well.
The complexity of this step depends on the methods:

\begin{itemize}
    \item OFM-$f_1$: local computation flops: $(nnz(\mathbf{A})2k + 4Nk^2 + Nk)/p$, communication cost: $6\alpha \log p + 4Nk\beta/\sqrt{p} + (2Nk\beta \log p) / p$;
    \item TriOFM-$f_1$: local computation flops: $(nnz(\mathbf{A})2k + 4Nk^2 + Nk)/p$, communication cost: $6\alpha \log p + 4Nk\beta/\sqrt{p} + (2Nk\beta \log p) / p$;
    \item OFM-$f_2$: local computation flops: $(nnz(\mathbf{A})2k + 8Nk^2 + 3Nk)/p$, communication cost: $8\alpha \log p + 4Nk\beta/\sqrt{p} + (4Nk\beta \log p) / p$;
    \item TriOFM-$f_2$: local computation flops: $(nnz(\mathbf{A})2k + 8Nk^2 + 3Nk)/p$, communication cost: $8\alpha \log p + 4Nk\beta/\sqrt{p} + (4Nk\beta \log p) / p$;
\end{itemize}

\subsection*{Complexity Analysis of Step 8 $\&$ 9}
The time complexity of steps 8 and 9 is independent of algorithms. The global summation of each column in step 8 is realized by local addition followed by the MPI\_Allreduce collective reducing and broadcasting $k$ words at each process in the same column. Thus, these two steps' local computation flops and communication costs are $k+7Nk/p$ and $2\alpha \log p + 2 \beta k \log p$, respectively.

\subsection*{Complexity Analysis of Step 10 $\&$ 11}
Given $\alpha^{(t)}$, step 11 only involves local computation, which takes $2Nk/p$ flops. The time complexity of step 10 varies among different algorithms. In general, step 10 involves multiple SpMMs, dot-products, local matrix-matrix multiplication, and trace summation. Some results have been evaluated in the previous steps, e.g., $\mathbf{A}\mathbf{X}$ and $\mathbf{X}^T\mathbf{X}$, so we do not need to repeat such redundant computations. According to Section 3.2, some intermediate matrices in the computation are the transpose of others. Thus, if they are available, plenty of time could be saved by using transposes. For example, $\mathbf{V}^T \mathbf{A} \mathbf{X}$ is the transpose of $\mathbf{X}^T \mathbf{A} \mathbf{V}$, so we only need to compute the latter. Furthermore, we assume finding the roots of a cubic polynomial via Cardano's formula takes constant $c$ flops. After careful calculation, we provide the complexity of steps 10 and 11 in each algorithm as follows:

\begin{itemize}
    \item OFM-$f_1$: local computation flops: $(nnz(\mathbf{A})2k+8Nk^2+2Nk)/p + 12k^3 + 4k^2 + 4k + c$, communication cost: $12\alpha \log p + 4Nk\beta / \sqrt{p} + (8Nk\beta \log p)/p$;
    \item TriOFM-$f_1$: local computation flops: $(nnz(\mathbf{A})2k+8Nk^2+2Nk)/p + 16k^3 + 6k^2 + (4+c)k$, communication cost: $12\alpha \log p + 4Nk\beta / \sqrt{p} + (8Nk\beta \log p)/p$;
    \item OFM-$f_2$: local computation flops: $(nnz(\mathbf{A})2k+8Nk^2+2Nk)/p + 18k^3 + 7k^2 + 4k + c$, communication cost: $12\alpha \log p + 4Nk\beta / \sqrt{p} + (8Nk\beta \log p)/p$;
    \item TriOFM-$f_2$: local computation flops: $(nnz(\mathbf{A})2k+8Nk^2+2Nk)/p + 32k^3 + 14k^2 + (4+c)k$, communication cost: $12\alpha \log p + 4Nk\beta / \sqrt{p} + (8Nk\beta \log p)/p$;
\end{itemize}

Adding up the complexity of steps 7-11, we get the total per-iteration complexity of every orthogonalization-free algorithm. Table \ref{tb:complexity} summarizes the asymptotic total complexity, assuming $k \ll N$, e.g., $k = \log N$.

\begin{table}[!htbp]
\caption{\textbf{Properties of matrices used in scalability test. The values under the ``load imb." column present the load imbalance in terms of the sparse matrix elements for 64 compute nodes (i.e., 8 $\times$ 8 2D partition).}}
\centering
\scalebox{1.0}{
\begin{tabular}{|c|c|c|c|c|}
\hline
Sparse Matrix & N & avg degree & nnz(A) & load imb.\\
\hline
LBOLBSV(SG)-1M & 1M & 48.4 & 48.4M & 1.15 \\
LBOHBSV(SG)-5M & 5M & 48.4 & 242.4M & 1.15 \\
LBOHBSV(SG)-20M & 20M & 48.5 & 970.1M & 1.15 \\
MAWI-Graph-1 & 18.6M & 3.0 & 56.6M & 6.76 \\
Graph500-scale24-ef16 & 16.8M & 31.5 & 529.4M & 5.47 \\
\hline

\end{tabular}
}
\label{tb:matrixproperties}
\end{table}

\begin{figure}[ht!]
  \begin{center}
    \begin{tabular}{c}
      \includegraphics[height=2.5in]{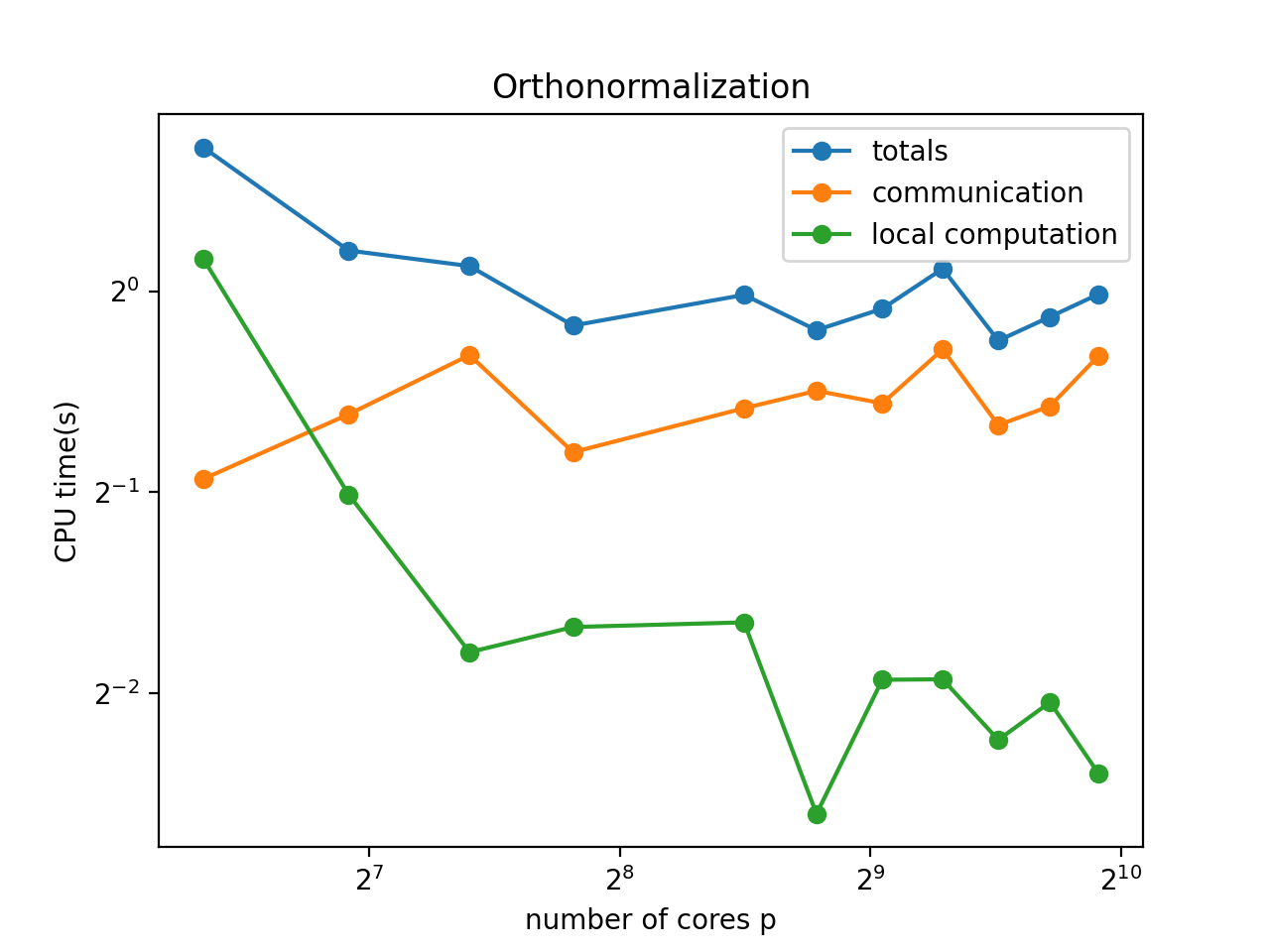} \\
    \end{tabular}
  \end{center}
\caption{Example scalability of parallel orthogonalization.}
\label{fig:orth-scaling}
\end{figure}

\begin{figure}[ht!]
  \begin{center}
    \begin{tabular}{ccc}
      \includegraphics[scale=0.30]{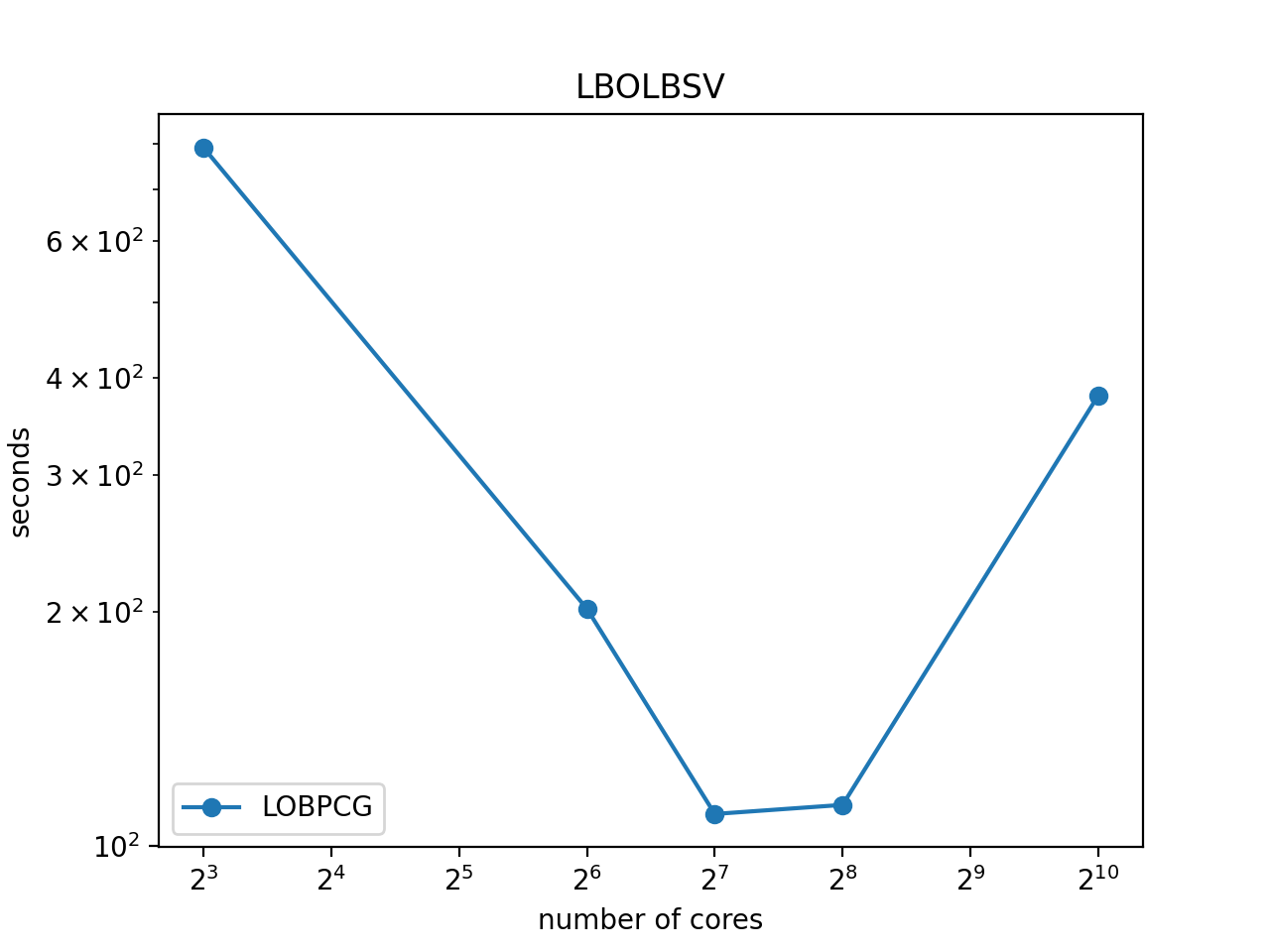} &
      \includegraphics[scale=0.30]{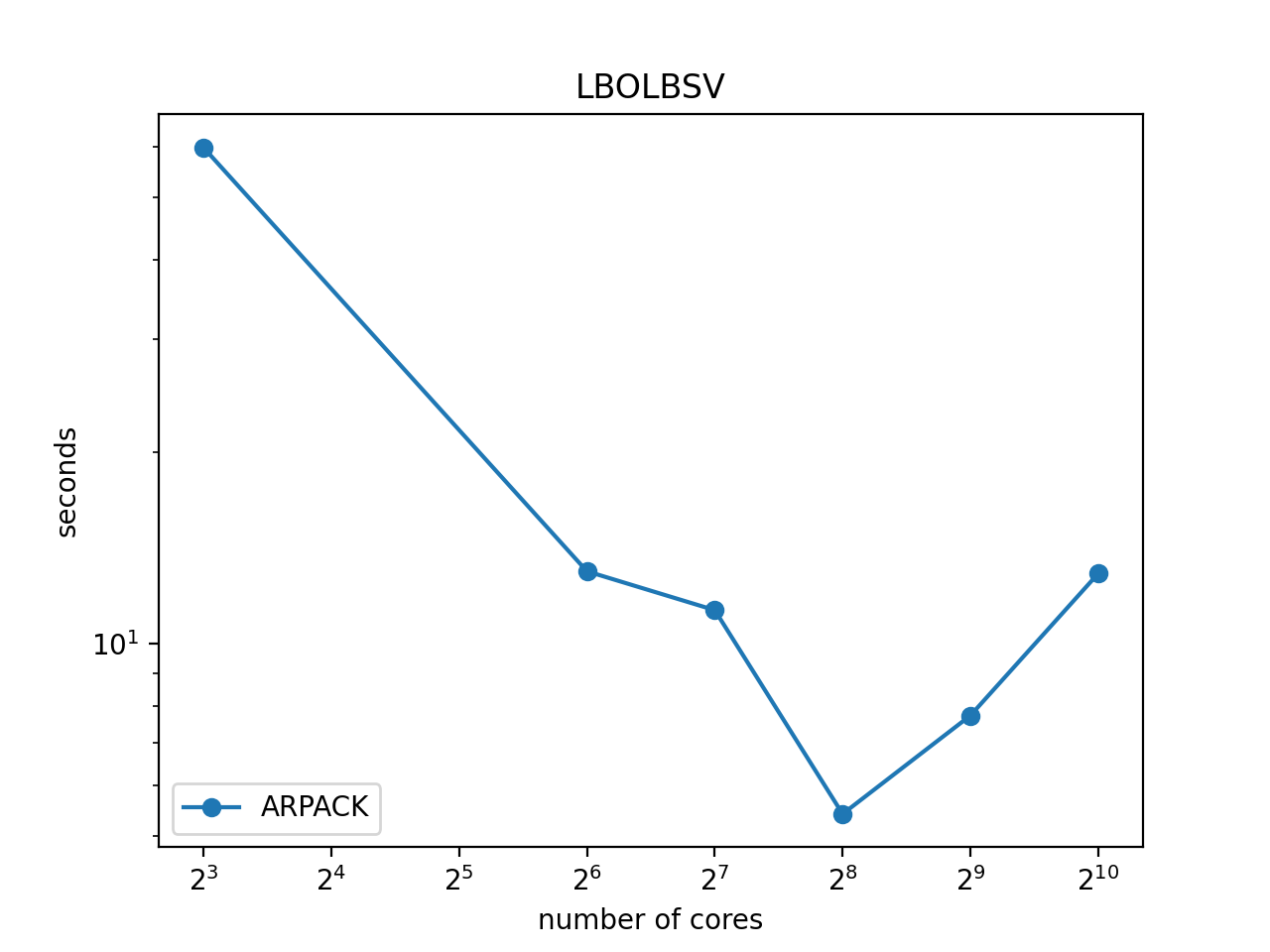} &
      \includegraphics[scale=0.30]{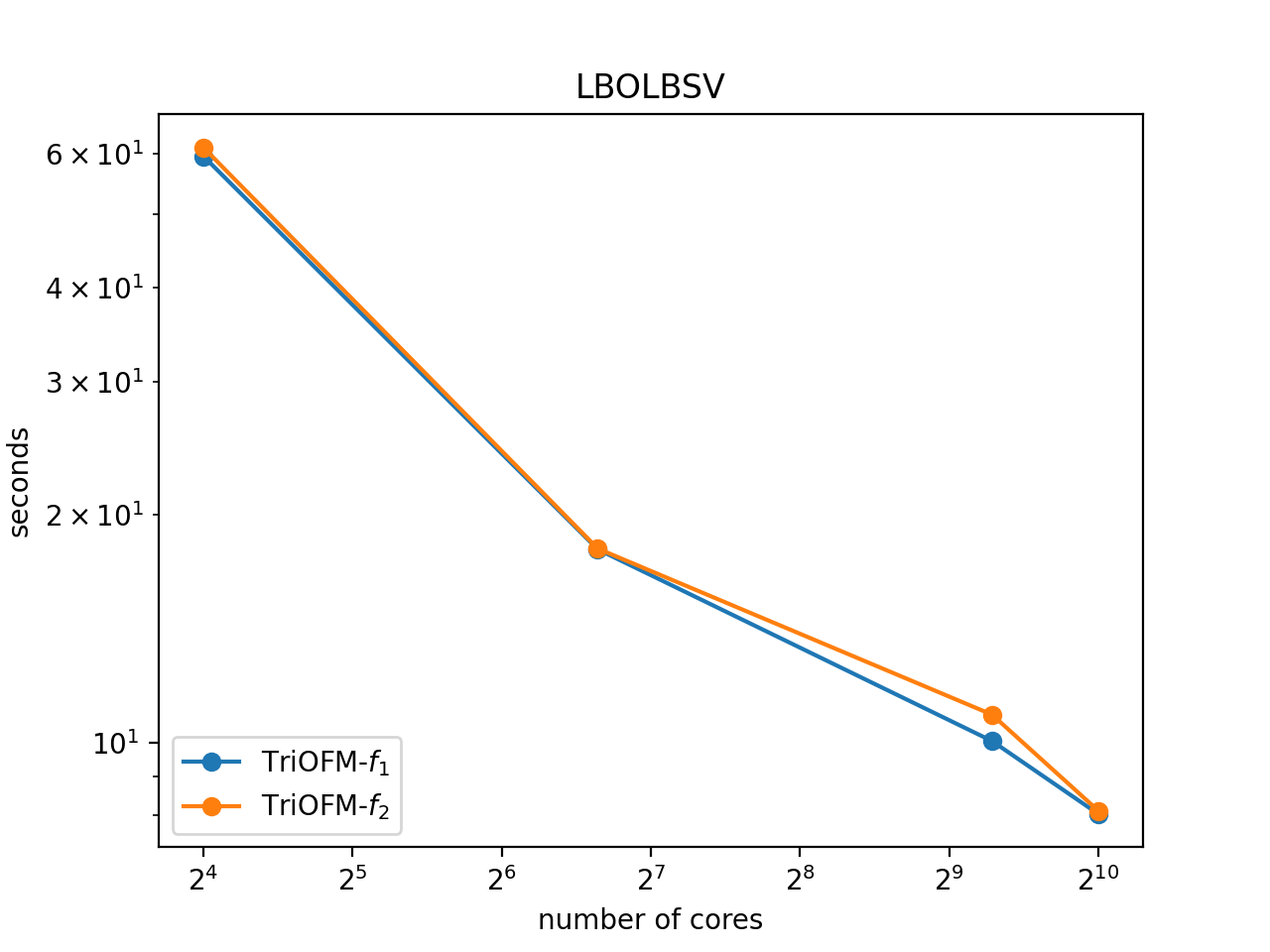} 
      
    \end{tabular}
  \end{center}
\caption{Scaling of ARPACK, LOBPCG, TriOFM-$f_1$, and TriOFM-$f_2$ on LBOLBSV(SG)-1M in Table \ref{tb:matrixproperties}, against the number of cores. The number of vectors $k = 64$. }
\label{fig:ofm-scalability-comp-cores}
\end{figure}

\begin{figure}[ht!]
  \begin{center}
    \begin{tabular}{cc}
      \includegraphics[scale=0.30]{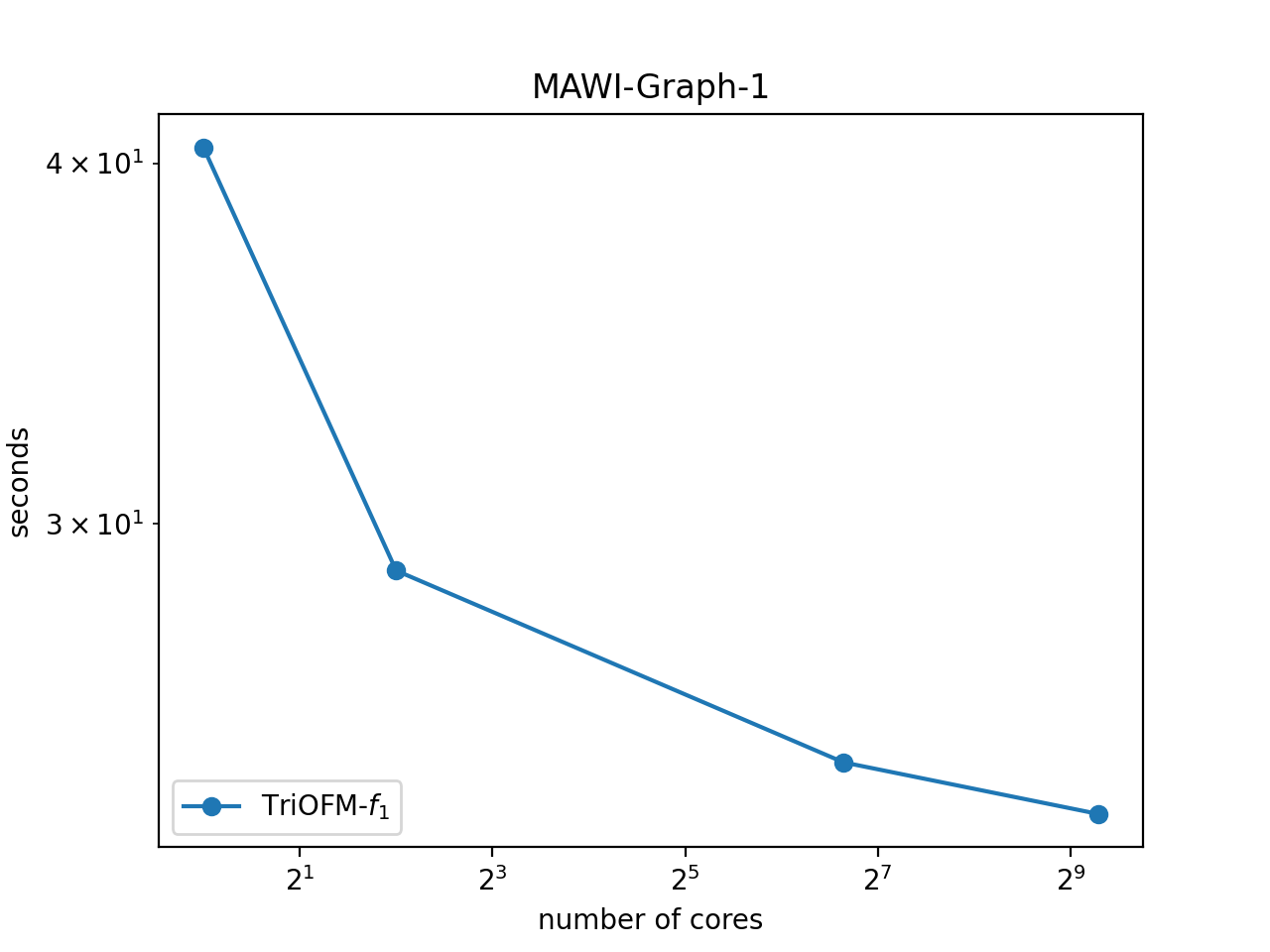} &  
      \includegraphics[scale=0.30]{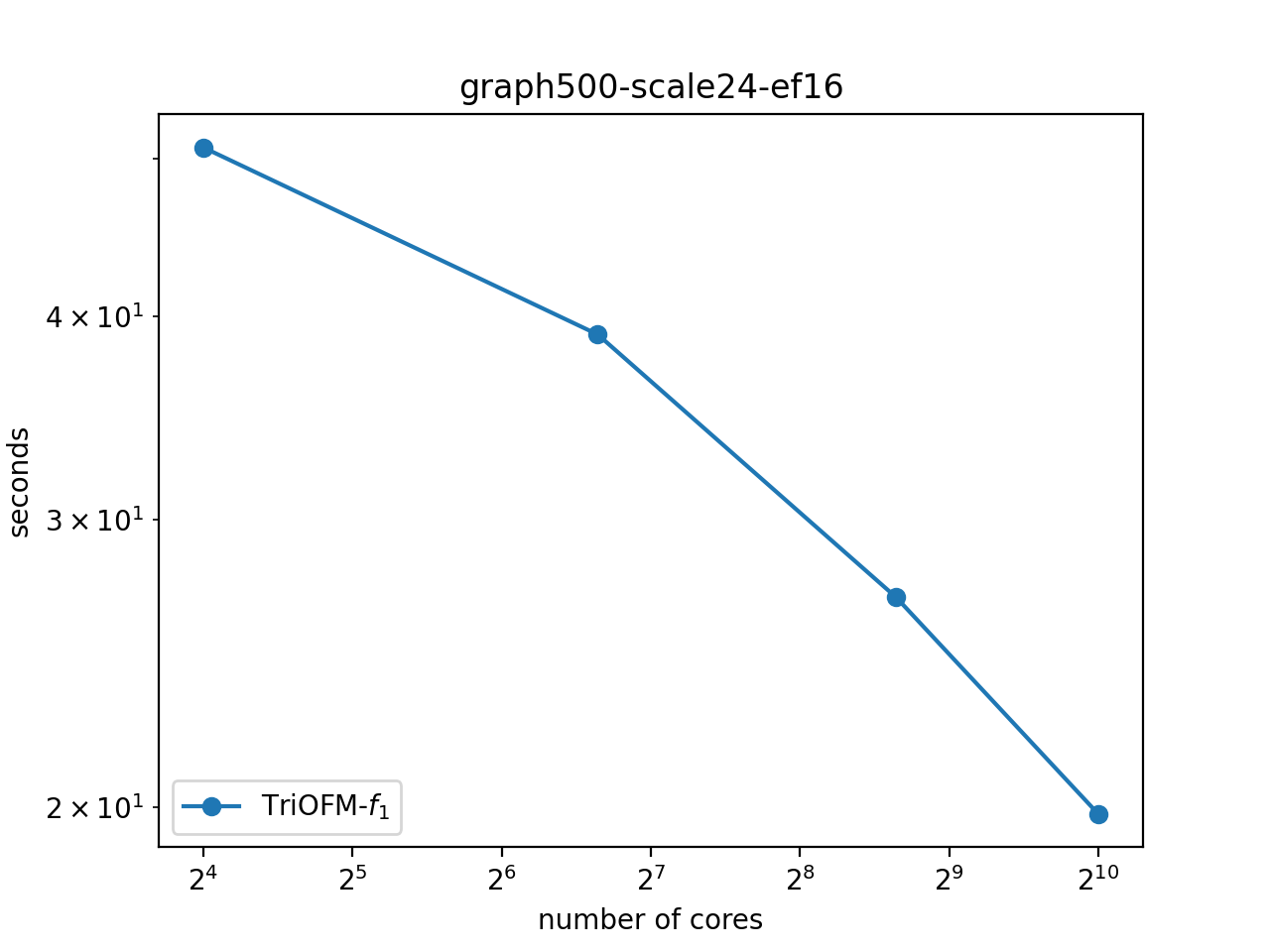}
      
    \end{tabular}
  \end{center}
\caption{Execution time in seconds that the orthogonalization-free methods use to compute $k = 8$ vectors on graphs MAWI-Graph-1 and Graph500-scale24-ef16 in Table \ref{tb:matrixproperties}, against the number of compute cores. }
\label{fig:ofm-scalability-cores}
\end{figure}

\begin{figure}[ht!]
  \begin{center}
    \begin{tabular}{c}
      \includegraphics[scale=0.35]{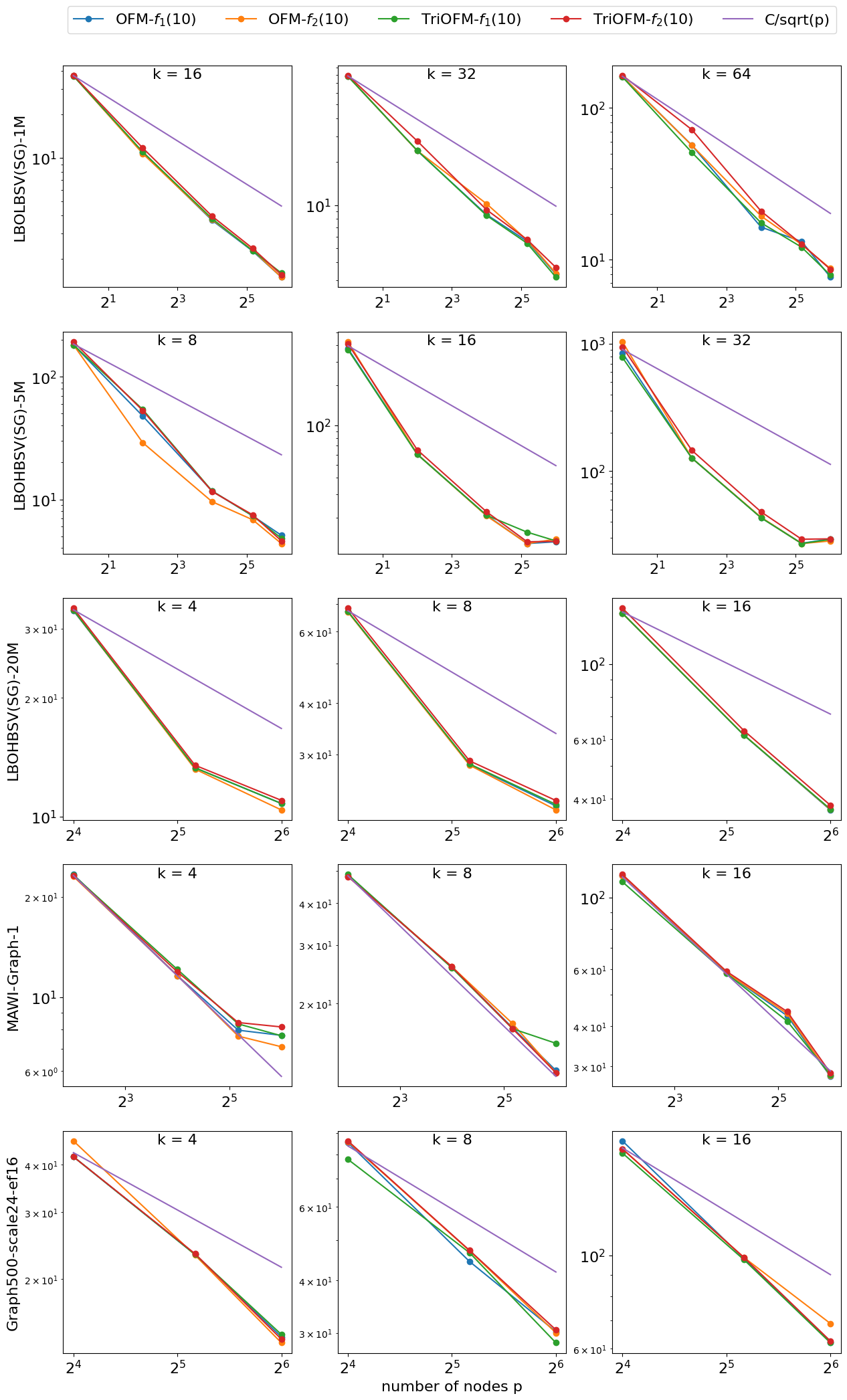} 
      
    \end{tabular}
  \end{center}
\caption{Execution time in seconds that the orthogonalization-free methods use to compute $k$ vectors on five graphs listed in Table \ref{tb:matrixproperties}, against the number of compute nodes. Only one process is used at one node. }
\label{fig:scalability}
\end{figure}

\section{Numerical Results}
This section presents the numerical results of five numerical experiments to demonstrate the effectiveness and scalability of the orthogonalization-free methods (OFMs). Our dataset consists of the synthetic graphs from the IEEE HPEC Graph Challenge. In the first experiment, we compare $\mathbf{U}_{k}$ and $\mathbf{U}_{k}\sqrt{-\Lambda_{k}}$ as features in spectral clustering and show no significant difference between them in clustering quality. In the second experiment, we compare directly using the results of OFMs and using the eigenvectors recovered from the results via the Rayleigh–Ritz method as features in spectral clustering. In the comparisons, orthogonalization does not improve clustering quality but takes additional overhead. In the third experiment, we compare the orthogonalization-free methods with other methods, i.e., ARPACK, LOBPCG, PI, and GSF, in both efficiency and clustering quality on synthetic static graphs. According to the results, OFMs show competitive clustering quality. In the fourth experiment, we compare the OFMs with ARPACK, LOBPCG, and PI on streaming graphs and show the advantage of OFMs in the scenario. GSF is excluded from comparisons on streaming graphs because it always takes particular Gaussian input and cannot take advantage of computed features in the scenario. The methods used in the first four tests are sequentially versions. In the final experiment, we discuss the scalability of each method and demonstrate the scalability advantages of the parallel orthogonalization-free methods.

All experiments are conducted on the Zaratan cluster operated by the University of Maryland. The cluster has 360 compute nodes, each with dual 2.45 GHz AMD 7763 64-core CPUs and HDR-100 (100 Gbit) Infiniband interconnects between the nodes, with storage and service nodes connected with full HDR (200 Gbit). The theoretical peak floating-point rate of the cluster is 3.5 Pflops. The code\footnote{https://github.com/qiyuanpang/DistributedLEVP.jl} available on Github is written in Julia 1.7.3 using OpenBLAS \cite{xianyi2012model} and Intel MKL \cite{wang2014intel} libraries for multithreading, and MPI.jl \cite{byrne2021mpi} back-ended by OpenMPI \cite{gabriel2004open} library for multiprocessing.

\subsection*{Dataset}
The synthetic graphs from Synthetic Data for the Streaming Graph Challenge: Stochastic Block Partition \footnote{http://graphchallenge.mit.edu} with known truth partitions fall into four categories: Low Block Overlap and Low Block Size Variation (the easiest task) (LBOLBSV), Low Block Overlap and High Block Size Variation (LBOHBSV), High Block Overlap and Low Block Size Variation (HBOLBSV), and High Block Overlap and High Block Size Variation (HBOHBSV). There are three sub-datasets in each category: Static Graphs (SG), Streaming Graphs - Edge Sampling (SGES), and Streaming Graphs - Snowball Sampling (SGSS). Hereafter, we use notations in the form of the category(sub-dataset)-size to represent a synthetic graph from the dataset. For example, LBOLBSV(SG)-1M represents the synthetic graph with 1 million graph nodes in the Static Graphs sub-dataset of the LBOLBSV category. Similarly, LBOLBSV(SGES)-1M represents the synthetic streaming graph with 1 million graph nodes in the Streaming Graphs - Edge Sampling sub-dataset of the LBOLBSV category. The only difference is that streaming graph(s) like LBOLBSV(SGES)-1M indeed consist of ten non-overlapping parts of a graph's edges.

\subsection*{Evaluation metrics}
Since the true partitions are known, we adopt two important external indexes to measure the similarity between the true partitions and the clustering results given by K-means clustering based on the features generated by the orthogonalization-free methods and other competitors. The two crucial indexes are Adjusted Rand Index (ARI) \cite{hubert1985comparing} and Normalized Mutual Information (NMI) \cite{danon2005comparing}. For ARI and NMI, values close to 0 indicate that the two clusterings are primarily independent, while values close to 1 indicate significant agreement. ARI is adjusted against chance, while NMI is not. To alleviate the randomness in Kmeans clustering, we repeat each experiment $10$ times to record the average indexes.

We also record the execution time in seconds for each method to compute features. Moreover, if we view all methods as eigensolver, we could measure the relative error (`relerr') as well: $\|\mathbf{B}\hat{\mathbf{U}}_{k} - \hat{\mathbf{U}}_{k}\hat{\mathbf{\Lambda}}_{k}\|_F / \|\hat{\mathbf{U}}_{k}\hat{\mathbf{\Lambda}}_{k}\|$, where $\mathbf{B}$ is the normalized similarity matrix with transition $2$ on the diagonal, and $\hat{\mathbf{\Lambda}}_{k}$ and $\hat{\mathbf{U}}_{k}$ are the corresponding eigenvalues and eigenvectors evaluated via the Rayleigh–Ritz method if necessary. We define $\mathbf{B}$ in this way to comply with ARPACK and LOBPCG, where we use $\mathbf{B}$ as the input matrix to guarantee positive definiteness. The input matrices for our methods, GSF and PI, are $\mathbf{A}$, $\mathbf{L}$, and the normalized similarity matrix. 

\subsection{Comparisons of features $\mathbf{U}_k$ and $\mathbf{U}_k \sqrt{-\Lambda_{k}}$}
In the first experiment, we compare the clustering quality of using features $\mathbf{U}_k$ and $\mathbf{U}_k \sqrt{-\Lambda_{k}}$ for spectral clustering. Note that the matrix $\mathbf{A}$ we use for OFMs and the normalized Laplacian $\mathbf{L}$ have the same eigenvectors $\mathbf{U}_k$. Eigenvectors $\mathbf{U}_k$ and eigenvalues $\Lambda_k$ of $\mathbf{A}$ are computed via LOBPCG without preconditioning under $0.1$ tolerance. Figure \ref{fig:compfeatures} summarizes the numerical comparisons and shows no significant difference in clustering quality between the two features. Therefore, we hypothesize that feature $\mathbf{U}_k \sqrt{-\Lambda_{k}}$ is also effective for spectral clustering. This hypothesis lays the foundation for the use of OFM-$f_1$ and TriOFM-$f_1$ in spectral clustering. Note that the features constructed by OFM-$f_2$ and TriOFM-$f_2$ are theoretically effective because they are isomorphic to the eigenvectors $\mathbf{U}_k$.

\subsection{Comparisons of OFMs with and without orthogonalization}
OFMs, followed by orthogonalization and the Rayleigh–Ritz method, can compute eigenvectors $\mathbf{U}_k$. After orthogonalizing the results given by OFMs, the Rayleigh–Ritz method \cite{leissa2005historical, ilanko2009comments, trefethen2022numerical} is applied to the orthogonal vectors to compute the Ritz pairs, which serve as the eigenvalues and eigenvectors of the original matrix $\mathbf{A}$. The Rayleigh–Ritz method, including one SpMM, several dot products, and an eigendecomposition of a $k\times k$ matrix, is cheap and scalable compared to the orthogonalization. In this experiment, we compare the clustering quality between two types of features--the direct results of OFMs and the eigenvectors evaluated by OFMs followed by orthogonalization and the Rayleigh–Ritz method. Figures \ref{fig:comporthf1} and \ref{fig:comporthf2} summarize the comparisons and show that orthogonalization and the Rayleigh–Ritz method do not improve clustering quality. Therefore, we conclude that orthogonalization is unnecessary when using OFMs for spectral clustering.

\subsection{Comparisons on Static Graphs}
In the third experiment, we compare the OFMs with GSF \cite{paratte2016fast}, PI-based methods in \cite{lin2010power}, ARPACK, and LOBPCG on static graphs. 

We first compare the approximators OFMs, GSF \cite{paratte2016fast}, and PI \cite{boutsidis2015spectral} because they do not evaluate eigenvectors. To make fair comparisons, we maintain the same number of sparse matrix-matrix multiplications for the orthogonalization-free methods and the PI-based method to compare clustering indexes and relative errors of computed eigenvectors. Therefore, we run our methods with $30$ iterations (e.g., `OFM-$f_1$($30$)') and the PI-based method `PI($61$)' with $61$ iterations. For GSF, we run a GSF `GSF($30,10$)' with a Jackson-Chebyshev polynomial of $30$ degree and $10$ iterations to estimate the $k$-th eigenvalue, as well as an ideal GSF `GSF($30,-$)' with the same Jackson-Chebyshev polynomial but with the exact smallest $k$-th eigenvalue given by LOBPCG. Figure \ref{fig:static-comp2} summarizes the comparison results, from where we can observe: 1) the orthogonalization-free methods and the PI-based method significantly outperform GSF in terms of clustering quality, execution time, and relative errors; 2) in terms of clustering quality, the orthogonalization-free methods are as good as the ideal GSF and outperform the PI-based method; 3) the four orthogonalization-free methods have similar performance.

Secondly, we compare the OFMs with ARPACK, LOBPCG without preconditioning, and LOBPCG with AMG preconditioning on static graphs. The tolerances set for ARPACK and LOBPCG are $0.1$ because OFMs with $30$ iterations achieve the same level of relative errors. The maximum iterations are set to be $200$ for both ARPACK and LOBPCG. The block size of LOBPCG is set to $16$. From Figure \ref{fig:static-comp4}, we observe that: 1) OFMs and LOBPCG with or without preconditioning achieve the top tier of clustering quality compared to ARPACK, but OFMs are more expensive than LOBPCG without preconditioning; 2) the AMG preconditioning seemingly does not accelerate LOBPCG on the static graphs.

\subsection{Comparisons on Streaming Graphs}
In the fourth experiment, we compare the OFMs, PI, ARPACK, and LOBPCG on streaming graphs. 
Since the edges of each graph are divided into ten disjoint parts coming in a streaming manner, in each stage, we add the coming part with the previous parts to build the current sub-graph and utilize the computed features for the previous sub-graph as the initial guess to the iterative methods for evaluating the features for the current sub-graph. Since ARPACK merely utilizes a one-dimensional starting vector, we set the starting vector to the computed eigenvector associated with the smallest eigenvalue if available. Larger block sizes in LOBPCG usually indicate more computation cost at each iteration due to the use of orthogonalization and the Rayleigh-Ritz method \cite{knyazev2017recent}. Though a larger block size in LOBPCG utilizes more previously computed eigenvectors as initial vectors for the current graph, there might be a balance between utilizing more initial vectors and total computation costs. Here, we set the block size in LOBPCG to $16$, the same as in the static graph scenario. The AMG preconditioning varies from different sub-graphs in the streaming graph scenarios, resulting in expensive computation costs. Therefore, we run LOBPCG without preconditioning in every stage for every sub-graph.
The input matrix to ARPACK and LOBPCG is the normalized Laplacian matrix with transition $2$ on the diagonals to ensure positive definiteness. The input matrices to an OFM and PI are matrix $\mathbf{A}$ and the normalized symmetric matrix, respectively. Note that GSF performs poorly in the static graph scenario and requires particular Gaussian inputs, so it cannot take advantage of any suitable initial vectors, which indicates that GSF is not competitive in streaming graph scenarios.

We run ARPACK and LOBPCG with prefixed tolerance $0.1$, the OFMs with $2$ iteration, and PI with $4$ iteration in each stage. The comparison results are summarized in Figures \ref{fig:streaming-comp-1M-e} and \ref{fig:streaming-comp-1M-s}. Surprisingly, OFMs and PI converge quickly on the streaming graphs and achieve the best clustering quality. 
OFMs and PI take similar computation costs on the data. OFM-$f_1$ and TriOFM-$f_1$ shows better clustering quality and relative errors compared to OFM-$f_2$ and TriOFM-$f_2$.
LOBPCG achieves the same clustering quality, but it is much more costly. ARPACK reaches the worst clustering quality and is also costly. Again, GSF performs poorly in the static graph scenario and cannot use computed features. Therefore, OFMs (especially OFM-$f_1$ and TriOFM-$f_1$) and PI have advantages over ARPACK, LOBPCG, and GSF in the streaming graph scenario.


\begin{figure}[ht!]
  \begin{center}
    \begin{tabular}{cc}
      \includegraphics[scale=0.50]{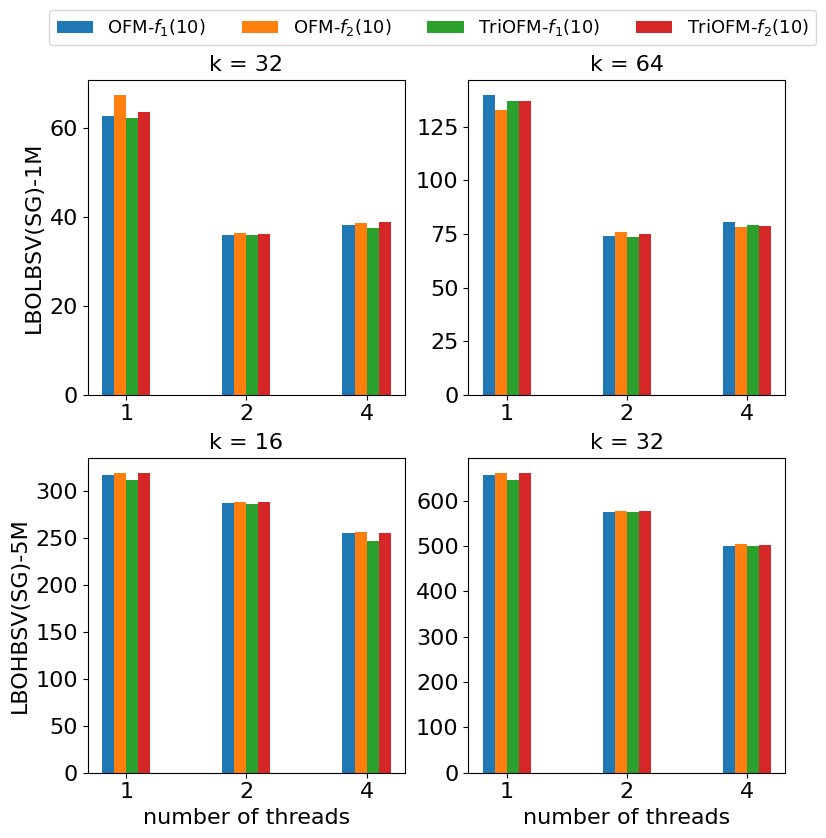} &
      
    \end{tabular}
  \end{center}
\caption{Execution time in seconds that the orthogonalization-free methods use to compute $k$ vectors on LBOLBSV(SG)-1M and LBOHBSV(SG)-5M, against the number of threads used in a process. Only one process is used. }
\label{fig:scalability2}
\end{figure}

\subsection{Scalability}
In the final experiment, we discuss the scalability of parallel versions of each method and demonstrate that OFMs are more scalable than other methods. We conduct the test on five graphs, including three from the Graph Challenge dataset, i.e., LBOLBSV(SG)-1M, LBOHBSV(SG)-5M, and LBOHBSV(SG)-20M, traffic data from the MAWI Project \cite{cho2000traffic} (MAWI-Graph-1), and synthetic graph data generated using the scalable Graph500 Kronecker generator \cite{kepner2018design} (Graph500-scale24-ef16). Various properties of these matrices used in the test are presented in Table \ref{tb:matrixproperties}. Load imbalance is defined as the ratio of the maximum number of nonzeros assigned to a process to the average number of nonzeros in each process:
\begin{equation}
    \dfrac{p * \max_{i,j}nnz(\mathbf{A}[i,j])}{nnz(\mathbf{A})}.
\end{equation}

In implementing four OFMs, the only difference lies in the "gradient" $ g$s, which include slightly different numbers of SpMMs and dot products in different OFMs. Consequently, all four OFMs should share the same scalability. Therefore, due to our limited computing resources, we only show the scalability of some OFMs in some scalability tests.
Figure \ref{fig:ofm-scalability-comp-cores} compares the scalability of OFMs, parallel ARPACK, and parallel LOBPCG without preconditioning on the graph LBOLBSV(SG)-1M up to $1024$ cores, and shows that OFMs are more scalable. The main reason is that parallel ARAPCK and LOBPCG use orthogonalization, which is not scalable in parallel computing environments. See Figure \ref{fig:orth-scaling} for an example. Since PI also employs orthogonalization to ensure efficient approximation to the eigenvectors in spectral clustering, an OFM is theoretically more scalable than PI, especially in the streaming scenario where orthogonalization takes a nonnegligible part in PI because both PI and the OFM converge fast. 

Figure \ref{fig:ofm-scalability-cores} further demonstrates the scalability of OFMs on two larger graphs MAWI-Graph-1 and Graph500-scale24-ef16, against numbers of cores. We also present the scalability of OFMs in Figure \ref{fig:scalability} against numbers of compute nodes in which only one process or core is used.  Figure \ref{fig:scalability} also demonstrates that the speedup is roughly $\sqrt{p}$ where $p$ is the number of compute nodes. All the aforementioned scalability results are for the MPI distributed memory setting.
Last, we also present the scalability of multithreading OFMs for shared memory in Figure \ref{fig:scalability2}. One may observe that the benefits of multithreading computing drop quickly as the number of threads increases. This performance entirely relies on the OpenBLAS and Intel MKL libraries and is a common phenomenon in multithreading computation, especially when the computation burden is relatively small.


\section{Conclusions}
We propose four orthogonalization-free methods for dimensionality reduction in spectral clustering by minimizing two objective functions. In contrast to current methods, including ARPACK, LOBPCG, the block Chebyshev-Davidson method, GSF, and PI, our methods do not require orthogonalization, which is known as not scalable in parallel computing environments. Theoretically, OFM-$f_2$ and TriOFM-$f_2$ construct features that are isomorphic to the eigenvectors $\mathbf{U}_k$ corresponding to the smallest eigenvalues of the symmetric normalized Laplacian of the graph. While OFM-$f_1$ and TriOFM-$f_1$ construct features that are isomorphic to the eigenvectors weighted by the square roots of eigenvalues, $\mathbf{U}_k \sqrt{-\Lambda_k}$. Numerical evidence shows that the feature $\mathbf{U}_k \sqrt{-\Lambda_k}$ is also effective in spectral clustering compared to $\mathbf{U}_k$, although more analytic analysis is needed in the future. Since the isomorphisms preserve Euclidean distance in Euclidean-distance-based Kmeans clustering for spectral clustering, orthogonalization is unnecessary in OFMs when used for spectral clustering. This is demonstrated by numerical results, which show that orthogonalization does not improve clustering quality in OFMs. Furthermore, numerical comparisons on synthetic graphs from the IEEE HPEC Graph Challenge show that the OFMs are competitive in the streaming graph scenario because OFMs can utilize all computed features for previous graphs and converge fast. In parallel computing environments, since orthogonalization is unnecessary, OFMs are more scalable than other methods and hence have the most potential to handle spectral clustering for large graphs, especially streaming graphs.

\section{Acknowledgments}
We thank Aleksey Urmanov from Oracle Lab at Oracle Corporation for the helpful discussion and comments. We thank Oracle Labs, Oracle Corporation, Austin, TX, for providing funding that supported research in the area of scalable spectral clustering and distributed eigensolvers.

\bibliographystyle{abbrv}
\bibliography{main}

\end{document}